\documentclass[a4paper,preprintnumbers,showpacs,onecolumn,superscriptaddress,nofootinbib,amsmath,amssymb,notitlepage]{revtex4-2}

\usepackage{mathtools,leftindex,tensor,mhchem}
\usepackage{stmaryrd}
\usepackage{booktabs}
\usepackage{siunitx}
\usepackage{enumitem}
\usepackage{times}
\usepackage{dcolumn}
\usepackage{mathrsfs}
\usepackage{comment}
\usepackage{hyperref}
\usepackage{amsmath,accents}
\usepackage{mathtools}
\usepackage{bbm}
\usepackage{bm}
\usepackage{amsfonts}
\usepackage{amssymb}
\usepackage{mathrsfs}
\usepackage[scr=rsfs,cal=boondox]{mathalfa}
\usepackage{wasysym}
\usepackage{graphicx}
\usepackage[]{subfigure}
\usepackage{filecontents}
\usepackage[dvipsnames]{xcolor}
\usepackage{bbold}
\usepackage[scr=rsfs,cal=boondox]{mathalfa}

\usepackage{stackengine}
\stackMath
\newcommand\tsup[2][2]{%
 \def\useanchorwidth{T}%
  \ifnum#1>1%
    \stackon[-1.3ex]{\tsup[\numexpr#1-1\relax]{#2}}{\mathchar"307E}%
  \else%
    \stackon[-1ex]{#2}{\mathchar"307E}%
  \fi%
}
\hypersetup{
    bookmarks=true,         
    pdftoolbar=true,        
    pdfmenubar=true,        
    pdffitwindow=true,     
    pdfstartview={FitH},     
    pdftitle={My title},     
    pdfauthor={author},      
    pdfsubject={Subject},    
    pdfcreator={Creator},    
    pdfproducer={Producer},  
    pdfkeywords={keyword1} 
    pdfnewwindow=true,       
    colorlinks=true ,        
    linkcolor=Blue  ,        
    citecolor=Blue,      
    filecolor=Blue,       
    urlcolor=Blue            
}

\newcommand{\ed}{\mathrm{d}}

\newcommand{\rr}{\mathrm{r}}

\newcommand{\Q}{\mathscr{Q}}

\newcommand{\oalpha}[1]{\accentset{\circ}{\alpha}}
\newcommand{\obf}[1]{\accentset{\circ}{\mathbf{f}}}
\newcommand{\boR}[1]{\accentset{\circ}{\mathbf{R}}}
\newcommand{\obF}[1]{\accentset{\circ}{\mathbf{F}}}
\newcommand{\obPi}[1]{\accentset{\circ}{\mathbf{\Pi}}}

\usepackage{scalerel}
\usepackage{tikz}
\usetikzlibrary{svg.path}

\definecolor{orcidlogocol}{HTML}{A6CE39}
\tikzset{
  orcidlogo/.pic={
    \fill[orcidlogocol] svg{M256,128c0,70.7-57.3,128-128,128C57.3,256,0,198.7,0,128C0,57.3,57.3,0,128,0C198.7,0,256,57.3,256,128z};
    \fill[white] svg{M86.3,186.2H70.9V79.1h15.4v48.4V186.2z}
                 svg{M108.9,79.1h41.6c39.6,0,57,28.3,57,53.6c0,27.5-21.5,53.6-56.8,53.6h-41.8V79.1z M124.3,172.4h24.5c34.9,0,42.9-26.5,42.9-39.7c0-21.5-13.7-39.7-43.7-39.7h-23.7V172.4z}
                 svg{M88.7,56.8c0,5.5-4.5,10.1-10.1,10.1c-5.6,0-10.1-4.6-10.1-10.1c0-5.6,4.5-10.1,10.1-10.1C84.2,46.7,88.7,51.3,88.7,56.8z};
  }
}

\newcommand\orcidicon[1]{\href{https://orcid.org/#1}{\mbox{\scalerel*{
\begin{tikzpicture}[yscale=-1,transform shape]
\pic{orcidlogo};
\end{tikzpicture}
}{|}}}}

\begin{document}


\title{Shadow analysis of an approximate rotating black hole solution with weakly coupled global monopole charge }

\author{Mohsen Fathi\orcidicon{0000-0002-1602-0722}}
\email{mohsen.fathi@ucentral.cl}
\affiliation{Centro de Investigaci\'{o}n en Ciencias del Espacio y F\'{i}sica Te\'{o}rica, Universidad Central de Chile, La Serena 1710164, Chile}


\begin{abstract}

We investigate the shadow properties of a rotating black hole with a weakly coupled global monopole charge, using a modified Newman-Janis algorithm. This study explores how this charge and rotational effects shape the black hole's shadow, causal structure, and ergoregions, with implications for distinguishing it from Kerr-like solutions. Analysis of null geodesics reveals observable features that may constrain the global monopole charge and weak coupling parameters within nonminimal gravity frameworks. Observational data from M87* and Sgr A* constrain the global monopole charge and coupling constant to $0 \leq \gamma \lesssim 0.036$ and $-0.2 \lesssim \alpha \leq 0$, respectively.

\bigskip

{\noindent{\textit{keywords}}: Black holes, nonminimal coupling, global monopole charge, Newman-Janis algorithm, shadow}\\

\noindent{PACS numbers}: 04.20.Fy, 04.20.Jb, 04.25.-g   

\end{abstract}

\maketitle


\section{Introduction}\label{sec:intro}

Black holes are among the most fascinating astrophysical phenomena predicted by general relativity (GR) and have been central to theoretical and observational research for decades. Observations of M87* \cite{Akiyama:2019_L1} and Sgr A* \cite{Akiyama:2022} by the Event Horizon Telescope (EHT) collaboration have intensified interest in black holes, allowing us to probe their properties beyond theoretical models. Despite GR’s successes at cosmic scales, it faces significant limitations when addressing quantum effects, topological defects, and certain early-universe phenomena. For instance, GR does not inherently include quantum-scale phenomena or high-energy structures, such as global monopoles, which are predicted by symmetry-breaking mechanisms in various field theories \cite{Vilenkin:1984ib, Barriola-Vilenkin:1989}. These monopoles, arising from phase transitions in the early universe, are characterized by a global monopole charge (GMC) that affects spacetime geometry. The presence of the GMC meaningfully alters black hole properties, influencing thermodynamic behavior, stability, and observable features such as photon orbits and shadow boundaries \cite{Achucarro:1999it, Barriola-Vilenkin:1989, Nucamendi:2001}. Studying black holes with a GMC thus offers a valuable connection between gravitational and particle physics, revealing possible limitations of GR and suggesting pathways for developing modified gravitational theories. In fact, topological defects like global monopoles are predicted by grand unified theories and arise from the Kibble mechanism, which links defect formation to the breaking of specific symmetry groups during cosmological phase transitions \cite{Kibble_1976, shellard_vilenkin}. A global monopole, resulting from the symmetry breaking of $\mathrm{SO}(3)$ to $\mathrm{U}(1)$, generates a unique gravitational effect that can be modeled by a Schwarzschild-like metric with an additional GMC term. This charge is connected to the symmetry-breaking energy scale and modifies spacetime by introducing a deficit angle, which affects the paths of test particles and the bending of light. Furthermore, the monopole’s core mass can be negative or negligible, leading to a repulsive gravitational effect that produces distinct features in black hole metrics. Furthermore, in modified gravity theories, the gravitational influence of global monopoles has been extensively studied, particularly in models that incorporate matter-curvature coupling \cite{CRomero, Liu_2009, Carames_2011, Carames_2017, Lambaga_2018, Nascimento_2019, gusmann_scattering_2021, GBgravity-2023}. A recent study \cite{carames_nonminimal_2023} in nonminimally coupled gravity showed that interactions between matter and geometry impart intriguing properties to global monopoles. In particular, the nonminimal coupling yields a positive core mass for the monopole and affects its internal structure, including core size. This study also examined the possibility of a nonminimal global monopole acting as an extra "hair" for a Schwarzschild black hole. By analyzing geodesic motion and gravitational light bending in the weak coupling regime, the study aimed to identify potential observational signatures of the GMC. This approach enables constraints on the nonminimal coupling and GMC parameters, providing insights into the symmetry-breaking scale. Recent work in Ref. \cite{FATHI2025169863} further investigated light propagation in the vicinity of this black hole, proposing observational limits on spacetime parameters.

Building on this interest in the impact of GMC on black hole properties, this paper examines the shadow of the rotating counterpart of this black hole. Analyzing the shadow of rotating black holes is essential, as it reveals observational signatures that can help differentiate various gravitational theories and enhance our understanding of spacetime geometry, especially in the presence of exotic features like topological defects. To obtain the rotating solution, we take the static metric introduced above as the seed and apply a modified version of the Newman-Janis algorithm (NJA) \cite{NewmanJanis:1965} to generate the rotating counterpart. This modified Newman-Janis algorithm (MNJA), as presented in Refs. \cite{Azreg:2014, Azreg:2014_1}, is particularly effective when extended to theories beyond GR, offering improved accuracy and stability in these contexts.

The paper is organized as follows: Sect. \ref{sec:overview} provides an overview of the static black hole solution with a weakly coupled GMC, covering its spacetime and causal structure. In Sect. \ref{sec:rotating}, we apply the MNJA step-by-step to derive the rotating counterpart of the static black hole. Sect. \ref{sec:CausalStructure} examines the causal structure of the resulting spacetime, including horizon formation under varying black hole parameters, as well as the ergoregions and their evolution in terms of the GMC and coupling constant. Sect. \ref{sec:H_J} introduces the Lagrangian formalism and the Hamilton-Jacobi equation to study light propagation in the black hole’s exterior geometry. Here, we derive equations of motion for null geodesics and identify conditions for spherical photon orbits, mapping photon regions outside the event horizon and demonstrating the effects of the GMC on these regions. This section also parametrizes the celestial plane for an observer at infinity, enabling shadow boundary plots for several cases and detailed analysis of their properties. To facilitate observational testing, Sect. \ref{sec:observables} introduces shadow observables and derives their mathematical expressions. We use these to constrain the GMC and coupling constant based on EHT data for M87* and Sgr A*. In Sect. \ref{sec:energy}, we calculate the black hole's evaporation rate, considering parameter constraints inferred from EHT data. Finally, we summarize our findings in Sect. \ref{sec:conclusion}. Throughout this paper, we adopt natural units with $G=c=1=M_{\rm{pl}}$, and use a sign convention of $(-,+,+,+)$, with primes denoting differentiation with respect to the radial coordinate.


\section{Static black hole solution with GMC in the weak coupling regime}\label{sec:overview}

The generalized theory of gravity incorporating nonminimal matter-curvature coupling is described by the action \cite{Bertolami:2007}
\begin{equation}
\mathcal{S} = \int \ed^4 x \sqrt{-g}\left\{\frac{1}{2}f_1(R)+\Bigl[1+\alpha f_2(R)\mathcal{L}_m\Bigr]\right\},
    \label{eq:action0}
\end{equation}
where $\mathcal{L}_m$ is the matter Lagrangian density, $f_1(R)$ and $f_2(R)$ are arbitrary functions of the Ricci scalar, and $\alpha$ represents the interaction strength between $f_2(R)$ and $\mathcal{L}_m$, effectively measuring the nonminimal coupling strength. In Ref. \cite{carames_nonminimal_2023}, the specific choices $f_1(R)=R/(8\pi)$ and $f_2(R)=R$ were considered. For these values, an upper limit of $|\alpha|< 5\times 10^{-12}\,\rm{m}^2$ was derived in Ref. \cite{PhysRevD.105.024020}, based on nuclear physics considerations at high densities \cite{PhysRevD.105.044048}. However, this limit is not applied in our study, as it pertains to environments with extreme densities not relevant here. Notably, setting $\alpha=0$ directly recovers GR for the above choices.

In Ref. \cite{carames_nonminimal_2023}, the matter Lagrangian density in Eq. \eqref{eq:action0} was chosen as 
\begin{equation}
\mathcal{L}_m = -\frac{1}{2} \partial_\mu \varphi^a \partial^\mu \varphi^a-\frac{1}{4}\lambda\bigl(\varphi^{a}\varphi^{a}-\eta^2\bigr)^2,
    \label{eq:Lm_0}
\end{equation}
which corresponds to symmetry breaking from $\rm{SO}(3)$ to $\rm{U}(1)$, giving rise to a global monopole. Here, $\lambda$ and $\eta$ denote the Higgs field's self-interaction constant and the symmetry-breaking energy scale, respectively. The Higgs field is represented by an isotriplet of scalar fields, $\varphi^a=\eta h(r) \hat{x}^a$, where $a=1,2,3$ and $x^a=\left\{\sin\theta \cos\phi,\sin\theta\sin\phi,\cos\theta\right\}$ in Schwarzschild coordinates $(t,r,\theta,\phi)$. The radial function $h(r)$ satisfies $h(0)=0$ and $h(\infty)=1$. This symmetry allows us to consider the spherically symmetric line element
\begin{equation}
\ed s^2=-B(r)\ed t^2+\frac{\ed r^2}{A(r)}+r^2\left(\ed\theta^2+\sin^2\theta\ed\phi^2\right),
    \label{eq:ds_0}
\end{equation}
for analyzing the field equations. The metric functions $B(r)$ and $A(r)$ were derived in Ref. \cite{carames_nonminimal_2023} by solving the field equations with this metric ansatz. Here, we specifically consider a weak matter-gravity coupling, where $\alpha f_2(R) = \alpha R < 1$. Thus, we focus on small values of $\alpha$, satisfying the condition $\alpha < l_0^2$, where $l_0$ is a characteristic length scale, approximately the black hole mass, $M$. In this weak coupling framework, the global monopole is treated as a GMC for a black hole of mass $M$, and Ref. \cite{carames_nonminimal_2023} provides solutions for $A(r)$ and $B(r)$ as:
\begin{eqnarray}
    && A(r) \approx 1-\gamma -\frac{2M}{r}+\frac{14\alpha\gamma}{r^2}-\frac{18\alpha\gamma M}{r^3},\label{eq:A(r)}\\
    && B(r) \approx 1-\gamma-\frac{2M}{r}+\frac{20 \alpha \gamma}{r^2}-\frac{30 \alpha \gamma M}{r^3},\label{eq:B(r)}
\end{eqnarray}
where $\gamma = 8\pi\eta^2$ represents the deficit solid angle. As a result, the spacetime resembles a conical cosmic string with a one-dimensional topological defect. Consequently, this spacetime is asymptotically finite with a conical singularity, yet does not reduce to Minkowski spacetime at infinity. The parameter $\alpha$ has dimensions of length squared, while $\gamma$ is dimensionless. In the absence of the GMC (i.e., $\gamma = 0$), the Schwarzschild black hole (SBH) solution is recovered. It is also important that $\eta^2 < 1$, ensuring that the symmetry-breaking scale remains below the Planck scale.

\begin{figure}[h]
    \centering
    \includegraphics[width=7 cm]{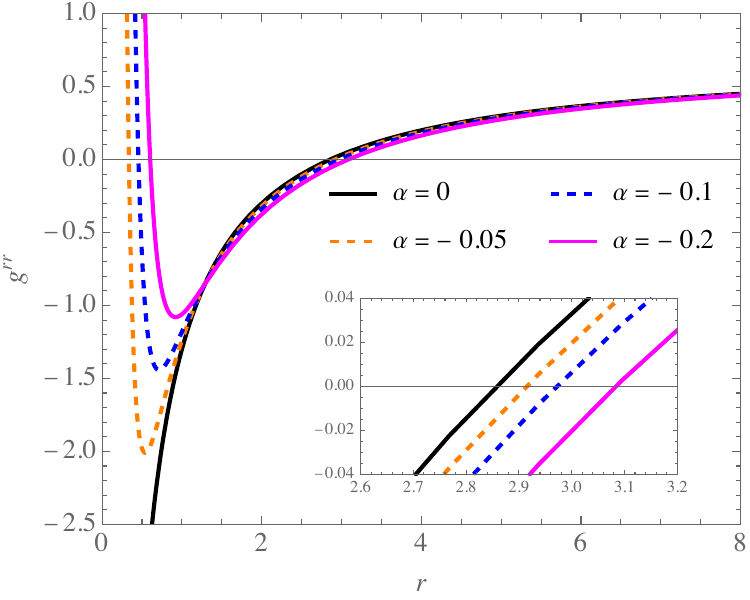} (a)\qquad
    \includegraphics[width=7 cm]{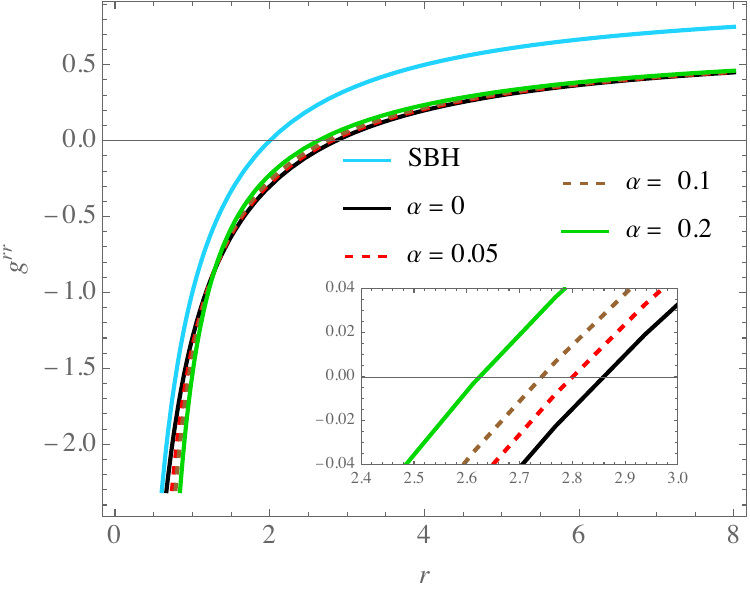} (b)
    \caption{The radial profile of $g^{rr}$ is plotted for $\gamma=0.3$. Diagrams correspond to (a) $\alpha\leq 0$ and (b) $\alpha\geq 0$, including the SBH case with $\gamma=1$. Length units are in terms of $M$.}
    \label{fig:A(r)-}
\end{figure}

The black hole's horizons occur where $g^{rr}=0$, yielding three solutions:
\begin{eqnarray}
    && \rr_1 = \frac{4}{1-\gamma} \sqrt{\frac{\zeta_2}{3}}\cos\left(\frac{1}{3}\arccos\left(\frac{2\zeta_3}{\zeta_2}\sqrt{\frac{3}{\zeta_2}}\,\right)-\frac{2\pi}{3}\right)+\frac{2M}{3(1-\gamma)},\label{eq:rr1}\\
    && \rr_2 =  \frac{4}{1-\gamma} \sqrt{\frac{\zeta_2}{3}}\cos\left(\frac{1}{3}\arccos\left(\frac{2\zeta_3}{\zeta_2}\sqrt{\frac{3}{\zeta_2}}\,\right)\right)+\frac{2M}{3(1-\gamma)} ,\label{eq:rr2}\\
    && \rr_3 = \frac{4}{1-\gamma} \sqrt{\frac{\zeta_2}{3}}\cos\left(\frac{1}{3}\arccos\left(\frac{2\zeta_3}{\zeta_2}\sqrt{\frac{3}{\zeta_2}}\,\right)-\frac{2\pi}{3}\right)+\frac{4M}{3(1-\gamma)} ,\label{eq:rr3}
\end{eqnarray}
where
\begin{eqnarray}
    \label{eq:zeta23}
&& \zeta_2=\frac{M^2}{3}-\frac{7\alpha}{2}\left(1-\gamma\right)\gamma,\\
&& \zeta_3 = \frac{M^2}{27}+\frac{M\alpha}{24}\left(27\gamma^2-40\gamma+13\right)\gamma.
\end{eqnarray}
In Fig. \ref{fig:A(r)-}, the behavior of $g^{rr}=A(r)$ is shown for various positive and negative values of $\alpha$ with fixed $\gamma$.
As illustrated, for $\alpha < 0$ (Fig. \ref{fig:A(r)-}(a)), there are two positive roots of $g^{rr} = 0$, indicating two horizons with $0 < \rr_1 < \rr_2$ and $\rr_3 < 0$. Here, $\rr_1$ represents the Cauchy horizon ($\rr_-$) and $\rr_2$ the event horizon ($\rr_+$). For $r > \rr_+$, observers remain time-like, while in the region $\rr_- < r < \rr_+$, no time-like observers exist. For $\alpha \geq 0$ (Fig. \ref{fig:A(r)-}(b)), only one real horizon exists, $\rr_+$, with a complex conjugate pair. The coupling impact is minimal in this case, but the topological defect encoded by $\gamma$ is significant, as shown by the deviation from the SBH profile. For this paper, we restrict our study to the outer communication domain, $r > \rr_+$. Additionally, the surface of infinite redshift, determined by $B(r) = 0$, provides insights into gravitational redshift effects. This surface has three solutions $\rr_4$, $\rr_5$, and $\rr_6$ that mirror the forms of $\rr_1$, $\rr_2$, and $\rr_3$, with $\zeta_2\rightarrow \tilde{\zeta}_2$ and $\zeta_3\rightarrow \tilde{\zeta}_3$, where
\begin{eqnarray}
    \label{eq:tzeta23}
     && \tilde{\zeta}_2=\frac{M^2}{3}-5\alpha\left(1-\gamma\right)\gamma,\\
     && \tilde{\zeta}_3 = \frac{M^3}{27}+\frac{5M\alpha}{24}  \left(9 \gamma ^2-14 \gamma +5\right) \gamma.
\end{eqnarray}
These conditions define the interior and exterior surfaces of infinite redshift, $\rr^{\rm{inf}}_{\rm{in}}=\rr_4$ and $\rr^{\rm{inf}}_{\rm{out}}=\rr_5$ for $\alpha < 0$, while for $\alpha > 0$, only one such surface, $\rr_5$, exists.

\section{The MNJA and the approximate rotating counterpart}
\label{sec:rotating}

In this section, we review the MNJA, which is applied for the generation of the stationary black hole minimally coupled with the GMC. The method is essentially based on the discussion provided in Refs. \cite{Azreg:2014, azreg-ainou_static_2014}.   

By considering the relation  
\begin{equation}
\ed t=\frac{\ed r}{\sqrt{A(r) B(r)}}+\ed u,
    \label{eq:dtdrdu}
\end{equation}
for the advanced null coordinate $u$, one can recast the line element \eqref{eq:ds_0} as
\begin{equation}
\ed s^2 = -B(r) \ed u^2-2\sqrt{\frac{B(r)}{A(r)}}\, \ed r\ed u + r^2\left(\ed\theta^2+\sin^2\theta\ed\phi^2\right).
    \label{eq:ds_null_1}
\end{equation}
This way, the null tetrad $\bm{e}_{(a)}=\left(\bm{l},\bm{n},\bm{m},\bar{\bm{m}}\right)$, satisfying the conditions $\bm{l}\cdot\bm{m}=\bm{n}\cdot\bm{n}=\bm{m}\cdot\bm{m}=\bm{l}\cdot\bm{m}=\bm{n}\cdot\bm{m}=0$ and $\bm{l}\cdot\bm{n}=-\bm{m}\cdot\bar{\bm{m}}=-1$, can generate the contravariant components of the metric tensor, as
\begin{equation}
g^{\mu\nu} = - l^\mu n^\nu-l^\nu n^\mu + m^\mu \bar{m}^\nu + m^\nu \bar{m}^\mu.
    \label{eq:gUU}
\end{equation}
By relating the above expressions to the line element \eqref{eq:ds_null_1}, we can choose
\begin{eqnarray}
    && l^\mu = \delta_r^\mu,\label{eq:lmu}\\
    && n^\mu = \sqrt{\frac{A(r)}{B(r)}}\,\delta_u^\mu-\frac{A(r)}{2}\delta_r^\mu,\label{eq:nmu}\\
    && m^\mu=\frac{1}{\sqrt{2}\,r}\left(\delta_\theta^\mu+\frac{\rm i}{\sin\theta}\,\delta_\phi^\mu\right).\label{eq:mmu}
\end{eqnarray}
Now by means of the complex transformations
\begin{eqnarray}
&& r\rightarrow r+\mathrm{i} a \cos\theta,\\
&& u\rightarrow u-\mathrm{i} a \cos\theta,
    \label{eq:ComplexTransf}
\end{eqnarray}
on the coordinates, one gets the null tetrad basis vectors in the complex coordinate system as
\begin{eqnarray}
&& l^{\mu}=\delta_{r}^\mu,\label{eq:ls}\\
&& n^{\mu}=\sqrt{\frac{\mathcal{A}}{\mathcal{B}}}\,\delta_{u}^{\mu}-\frac{\mathcal{A}}{2}\,\delta_{r}^\mu,\label{eq:ns}\\
&& m^{\mu}=\frac{1}{\sqrt{2\Psi}}\left[\delta_{\theta}^\mu+\mathrm{i} a\sin\theta\left(\delta_{u}^\mu-\delta_{r}^\mu\right)+\frac{\rm i}{\sin\theta}\,\delta_{\phi}^\mu\right],\label{eq:ms}
\end{eqnarray}
in which we have performed the replacements
\begin{subequations}
    \begin{align}
        & A(r)\rightarrow \mathcal{A}(r,\theta,a),\\
        & B(r)\rightarrow \mathcal{B}(r,\theta,a),\\
        & r^2\rightarrow \Psi(r,\theta,a),
    \end{align}
    \label{eq:mathcalABPsi}
\end{subequations}
in the expressions \eqref{eq:lmu}--\eqref{eq:nmu}. It is then straightforward to recover the seed static spherically symmetric spacetime metric \eqref{eq:ds_null_1}, by imposing 
\begin{eqnarray}
&& \lim_{a\rightarrow0}\mathcal{A}(r,\theta,a)=A(r),\label{eq:limmA}\\
&& \lim_{a\rightarrow0}\mathcal{B}(r,\theta,a)=B(r),\label{eq:limmB}\\
&& \lim_{a\rightarrow0}\Psi(r,\theta,a)=r^{2}.\label{eq:limPsi}
    \label{eq:a0_recover}
\end{eqnarray}
In the original NJA, as presented in Ref. \cite{newman_note_1965}, the functions $\mathcal{A}$, $\mathcal{B}$, and $\Psi$ are derived through the complexification of the radial coordinate. However, in the MNJA, this complexification is not imposed. Instead, these functions are determined under specific criteria, which are governed by a particular form of the energy-momentum tensor. By applying the expressions in \eqref{eq:ls}--\eqref{eq:ms} within the contravariant metric \eqref{eq:gUU}, we obtain
\begin{equation}
g^{\mu\nu} = 
    \renewcommand{\arraystretch}{1.8} 
    \begin{pmatrix}
        \displaystyle \frac{a^2 \sin^2\theta}{\Psi} & 
        \displaystyle -\frac{1}{2}\left[\sqrt{\frac{\mathcal{A}}{\mathcal{B}}} + \frac{a^2 \sin^2\theta}{\Psi}\right] & 0 & 
        \displaystyle \frac{a}{2\Psi}\\
        
        \displaystyle -\frac{1}{2}\left[\sqrt{\frac{\mathcal{A}}{\mathcal{B}}} + \frac{a^2 \sin^2\theta}{\Psi}\right]  & 
        \displaystyle \mathcal{A} + \frac{a^2 \sin^2\theta}{\Psi} & 0 & 
        \displaystyle -\frac{a}{2\Psi}\\
        
        0 & 0 & \displaystyle \frac{1}{\Psi} & 0\\
        
        \displaystyle \frac{a}{2\Psi} & 
        \displaystyle -\frac{a}{2\Psi} & 0 & 
        \displaystyle \frac{1}{\Psi \sin^2\theta}\\
    \end{pmatrix}.
    \label{eq:gUU_complex}
\end{equation}
Accordingly, the line element in the advanced null coordinates is rewritten as
\begin{multline}
\ed s^{2} = -\mathcal{B} \ed u^{2}-2\sqrt{\frac{\mathcal{B}}{\mathcal{A}}}\,\ed u\ed r + \Psi \ed\theta^2-2a\sin^2\theta\left(\sqrt{\frac{\mathcal{A}}{\mathcal{B}}}-\mathcal{A}\right)\ed u\ed \phi+2a\sin^2\theta\sqrt{\frac{\mathcal{A}}{\mathcal{B}}}\,\ed r\ed\phi \\
+\sin^2\theta\left[
\Psi+a^2\sin^2\theta\left(2\sqrt{\frac{\mathcal{A}}{\mathcal{B}}}-\mathcal{A}\right)
\right]\ed\phi^2.
    \label{eq:ds_null_2}
\end{multline}
Now to cast the above line element in the Boyer-Lindquist coordinates, we first perform the global coordinate transformations 
\begin{eqnarray}
    && \ed u=\ed t+\chi_1(r)\ed r,\label{eq:du_1}\\
    && \ed\phi = \ed\phi+\chi_2(r)\ed r,\label{eq:dphi_1}
\end{eqnarray}
which yields $g_{r\phi} = 0 = g_{tr}$. {However, this condition is not always satisfied in the original NJA, as the functions $\chi_{1,2}$ may also depend on the polar coordinate $\theta$. In such cases, a complete differential on the right-hand side of Eqs. \eqref{eq:du_1} and \eqref{eq:dphi_1} cannot be obtained, and as a result, the coordinates $u$ and $\phi$ do not exist. Consequently, for rotating solutions derived using the NJA, the separability of the Hamilton-Jacobi equation for null geodesics holds only when $\chi_{1,2}$ are functions of $r$ alone. This restriction makes it possible to express the metric in Boyer-Lindquist coordinates, as discussed in Ref. \cite{Shaikh:2020}. The limitation of the original NJA arises from the fact that the functions $\mathcal{A}$, $\mathcal{B}$, and $\Psi$ are determined by the complexification of the radial coordinate. In contrast, in the MNJA, these functions are not identified through complexification.} Nevertheless, based on the method proposed in Ref. \cite{Azreg:2014}, by fixing
\begin{eqnarray}
&& \chi_1(r) = -\frac{K(r)+a^2}{\Delta(r)},\label{eq:chi_1}\\
&& \chi_2(r) = - \frac{a}{\Delta(r)},\label{eq:chi_2}
\end{eqnarray}
where $\Delta(r)=r^2 A(r)+a^2$, and 
\begin{equation}
K(r) = r^2\sqrt{\frac{A(r)}{B(r)}},
    \label{eq:K}
\end{equation}
we can always recast the line element \eqref{eq:ds_null_2} in Boyer-Lindquist coordinates, by considering the definitions
\begin{eqnarray}
    && \mathcal{A}(r,\theta,a) = \frac{\Delta -a^2\sin^2\theta}{\Psi},\label{eq:mA}\\
    && \mathcal{B}(r,\theta,a) = \frac{\Delta-a^2\sin^2\theta}{\rho^4(r,\theta)}\Psi,\label{eq:mB}
\end{eqnarray}
where $\rho^2(r,\theta) = K(r) + a^2 \cos^2\theta$. It can be verified that the above equations satisfy the conditions \eqref{eq:limmA}--\eqref{eq:limPsi}. Now, the expressions \eqref{eq:mA} and \eqref{eq:mB}, together with Eq. \eqref{eq:ds_null_2}, lead to the new Kerr-like line element
\begin{multline}
\ed s^2 = \frac{\Psi}{\rho^2}\Biggl\{- \left(\frac{\Delta-a^2\sin^2\theta}{\rho^2}\right)\ed t^2 + \frac{\rho^2}{\Delta}\,\ed r^2+\rho^2\,\ed\theta^2+2a\sin^2\theta\left(\frac{\Delta-a^2-K}{\rho^2}\right)\ed t\ed\phi\\
+\sin^2\theta\left[
\rho^2-a^2\sin^2\theta\left(
\frac{\Delta-a^2\left(2-\sin^2\theta\right)-2K}{\rho^2}
\right)
\right]\ed\phi^2
\Biggr\},
 \label{eq:ds_kerr_like}
\end{multline}
in the Boyer-Lindquist coordinates, with $a = J/M$, where $J$ is the black hole's angular momentum, the spin parameter $a$ is defined. Note that the function $\Psi(r,\theta,a)$ remains unknown at this stage. However, there are criteria through which this function may be determined. For instance, as stated in Ref. \cite{azreg-ainou_static_2014}, the function $\Psi$ satisfies the nonlinear differential equation
\begin{equation}
\left(K+a^2 y^2\right)^2\bigl(3\partial_r\Psi\partial_{yy}\Psi-2\Psi\partial_{ryy}\Psi\bigr) = 3a^2\partial_r K\Psi^2,
    \label{eq:Psi_cond_1}
\end{equation}
where $y \equiv \cos\theta$. The above condition corresponds to the vanishing of the $r\theta$-component of the Einstein tensor, when the source is an imperfect fluid rotating about the $z$-axis. In the special case of $K = r^2$ (as in the cases of $g^{rr} = -g_{tt}$ for the seed static spherically symmetric line elements), it can be verified that $\Psi = r^2 + a^2 y^2$ is a possible solution to Eq. \eqref{eq:Psi_cond_1}.
Hence, the Kerr black hole (KBH) can be obtained by letting $\gamma = 0$ in the metric functions $A(r)$ and $B(r)$. However, since $A(r) \neq B(r)$ in the line element \eqref{eq:ds_0}, the function $\Psi$ cannot be easily fixed. It is straightforward to check that for $a = 0$, the stationary metric \eqref{eq:ds_kerr_like} reduces to the static one \eqref{eq:ds_0}, in the case where $\Psi = r^2$. On the other hand, it can be observed that the function $\Psi$ appears as a conformal factor in the stationary line element \eqref{eq:ds_kerr_like}, and thus, it does not affect the causal structure of the spacetime nor the evolution of the null geodesics in the black hole's exterior. 

However, {it is important to emphasize that, while the MNJA, as summarized here and originally developed in Ref. \cite{Azreg:2014}, does not exhibit any evident pathologies, certain inconsistencies may arise depending on how the method is applied to generate rotating solutions \cite{Rodrigues:2017}. Despite this, it should be noted that the global monopole in the matter Lagrangian of the theory under investigation is presumed to be minimally coupled to gravity. As a result, deviations from GR are expected to be minimal. Nevertheless, given the complexity involved in deriving the final rotating solution through the aforementioned steps, it must be acknowledged that the resulting solution is merely an approximation of the true rotating solution. The rotating spacetime described by the line element \eqref{eq:ds_kerr_like} contains the GMC, thus representing the exterior spacetime of a rotating monopole black hole (RMBH). Moreover, as discussed in Ref. \cite{junior_spinning_2020}, the Hamilton-Jacobi equation for null geodesics traveling in the resulting stationary spacetime obtained from the MNJA is always separable. Consequently, the MNJA has been widely employed in the literature to generate rotating solutions in Boyer–Lindquist-like coordinates from static black holes in both general relativity and alternative gravity theories (see, for example, Refs. \cite{xu_kerr-newman-ads_2017,toshmatov_rotating_2017,toshmatov_generic_2017, kumar_rotating_2018,azreg-ainou_rotating_2019,contreras_black_2019,haroon_shadow_2020,kumar_rotating_2020,contreras_black_2020,jusufi_rotating_2020,Shaikh:2020,liu_shadow_2020,junior_spinning_2020,fathi_ergosphere_2021,fathi_spherical_2023,fathi_spherical_particle_2023,zahid_shadow_2024,raza_shadow_2024}).}

Beginning in the next section, we will begin studying the rotating black hole described above, focusing on the causal structure of the spacetime and the behavior of photons in its vicinity.

\section{Horizons and the causal structure}\label{sec:CausalStructure}

\begin{figure}[h]
    \centering
    \includegraphics[width=7cm]{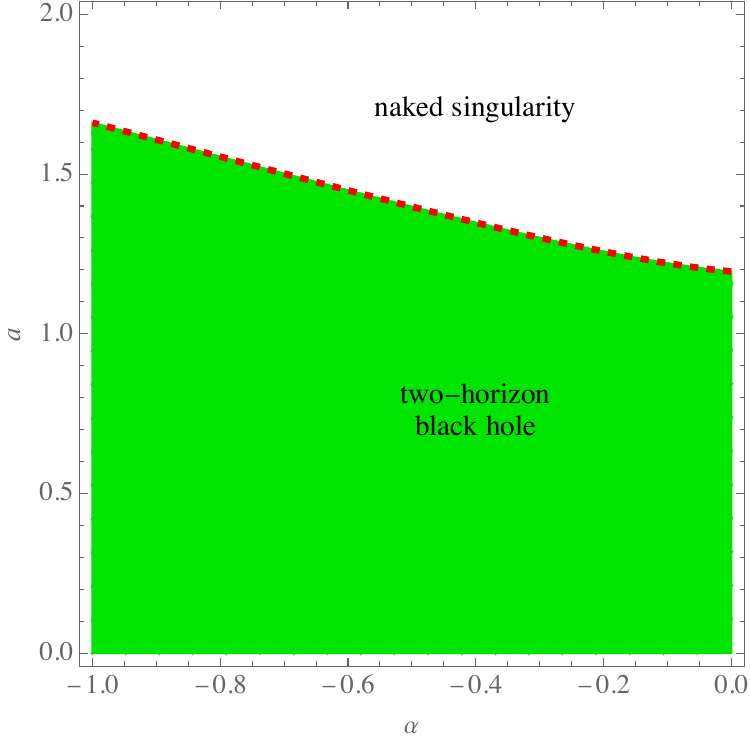} (a)\qquad\qquad
    \includegraphics[width=7cm]{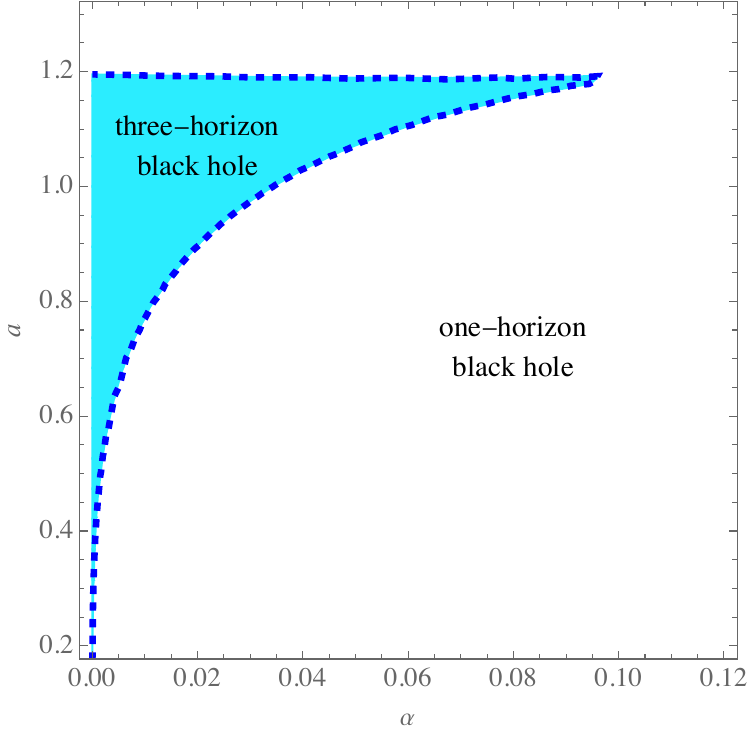} (b)
    \caption{The region plots of $\mathcal{D}_{\Delta} \geq 0$ (the colored areas) for $\gamma = 0.3$, showing the changes of $a$ versus $\alpha$ within this criterion. Panel (a) corresponds to the case of $\alpha \leq 0$, for which the region where $\mathcal{D}_{\Delta} > 0$ corresponds to two positive and one negative real roots. Hence, the black hole has two horizons. The dashed red curve corresponds to the extremal black hole (EBH), for which $\mathcal{D}_\Delta = 0$, and the black hole has only one horizon, as the positive roots become degenerate. Beyond this line, there will be only one negative real root and two complex conjugate ones, leading to the appearance of a naked singularity. In panel (b), where the $\alpha > 0$ cases are considered, the region where $\mathcal{D}_{\Delta} > 0$ again has three real roots, which are now all positive. Therefore, the black hole will have three horizons in this domain. The blue dashed line in this case again corresponds to $\mathcal{D}_{\Delta} = 0$, providing a black hole with two horizons, as two of the aforementioned real roots merge. This represents a different kind of extremality, leaving a black hole with an outer and an inner Cauchy horizon. Beyond this limit, where $\mathcal{D}_{\Delta} < 0$, the black hole has only one horizon.
    The unit of length along the vertical axis is $M$.}
    \label{fig:DiscDelta}
\end{figure}

\begin{figure}[h]
    \centering
    \includegraphics[width=7cm]{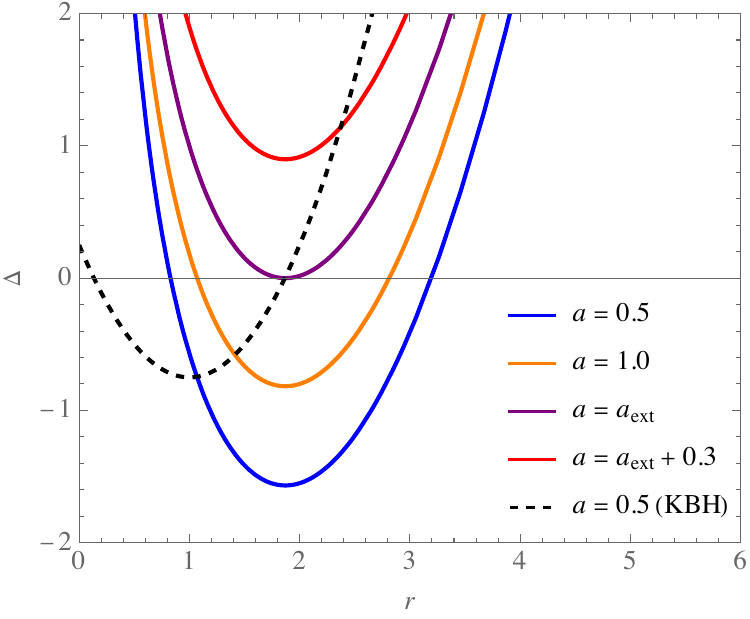} (a)\qquad\qquad
    \includegraphics[width=7cm]{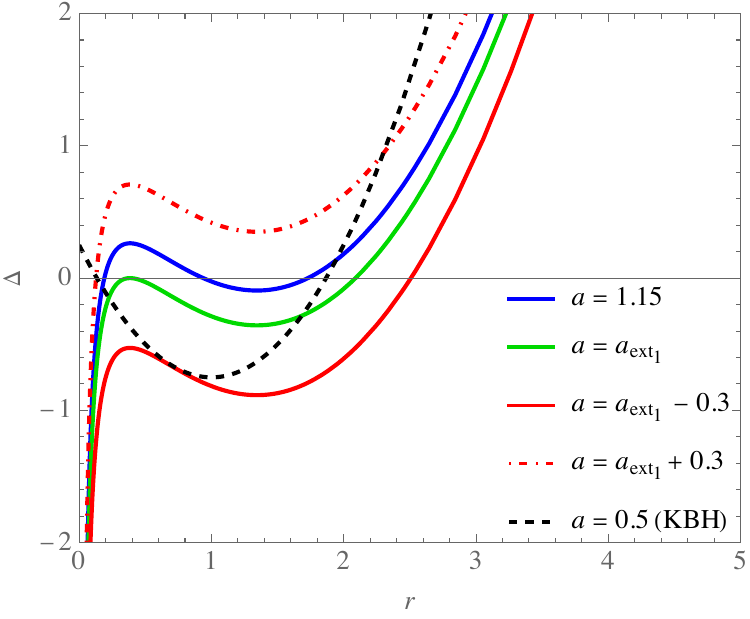} (b)
    \caption{The radial profiles of the $\Delta$-function plotted for $\gamma = 0.3$ and for (a) $\alpha = -0.4$, where $a_{\rm{ext}} = 1.348$, and (b) $\alpha = 0.04$, where $a_{\rm{ext}_1} = 1.029$ and $a_{\rm{ext}_2} = 1.190$. The dashed black curve represents the profile for the KBH, i.e., for $\gamma = 0$, with a spin parameter of $a = 0.5$. The units of length along the axes are in terms of the mass $M$.
}
    \label{fig:Delta}
\end{figure}

\begin{figure}[h]
    \centering
    \includegraphics[width=7.5cm]{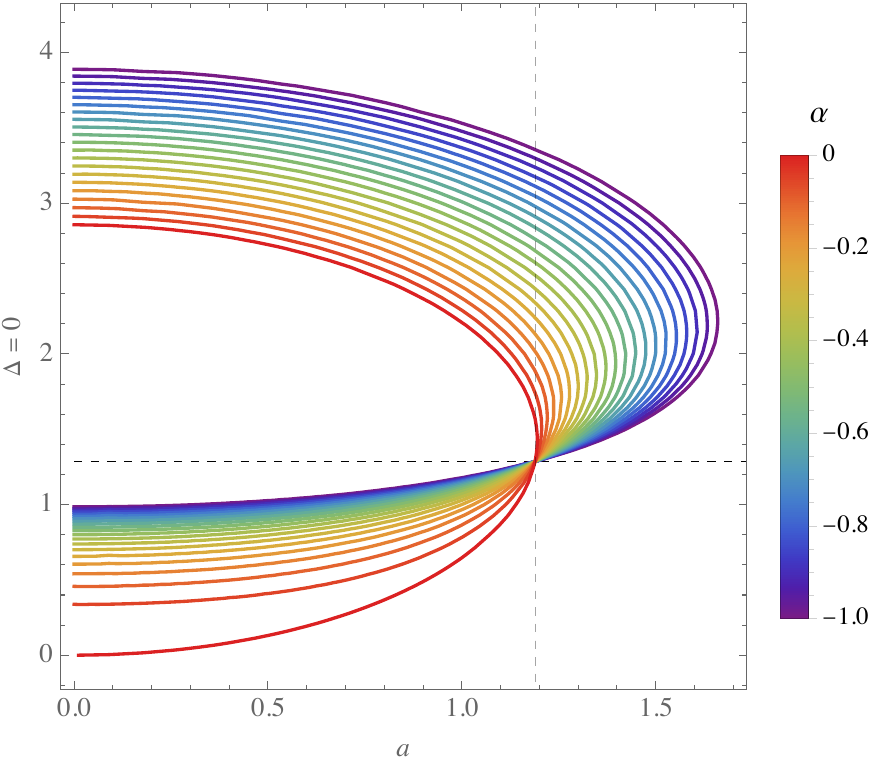} (a)\qquad\qquad
    \includegraphics[width=7.5cm]{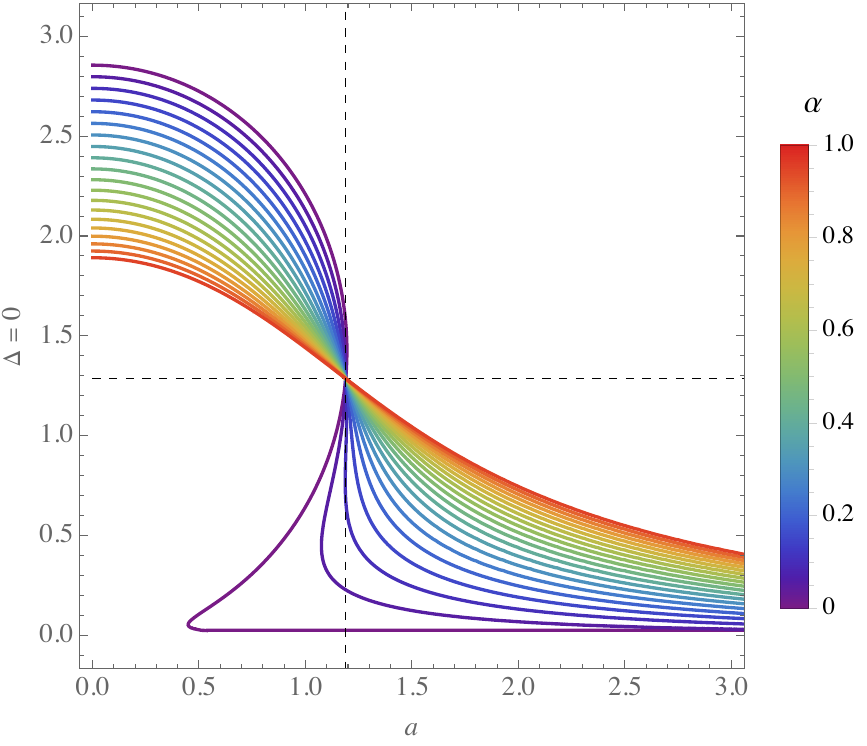} (b)
    \caption{The behavior of the equation $\Delta = 0$ with respect to changes in the spin parameter plotted for $\gamma = 0.3$ and different values of the $\alpha$-parameter. Panels (a) and (b) correspond to $\alpha \leq 0$ and $\alpha > 0$, respectively. The dashed lines indicate a point that is common to all the curves, characterized by $a = 1.189$ and $r = 1.286$ in this case. The unit of length along the axes is in terms of the mass $M$.}
    \label{fig:Delta=0}
\end{figure}

The black hole horizons are identified by the polynomial equation $\Delta = 0$, whose discriminant is given by
\begin{multline}
\mathcal{D}_{\Delta}\equiv\mathrm{Disc}(\Delta=0)=-10976 \left(1-\gamma\right) \gamma ^3\alpha^3+4\Bigl[\Bigl(81  \left(26-27 \gamma \right)\gamma+277\Bigr) M^2-588  \left(1-\gamma\right)a^2\Bigr]\gamma^2\alpha^2\\
+  \Bigl[ \left(95-81 \gamma \right)  a^2 M^2-72 M^4-21  \left(1-\gamma\right)a^4\Bigr]\gamma \alpha 
+4 a^4 M^2-4 \left(1-\gamma\right) a^6.
    \label{eq:disc_cubic}
\end{multline}
In Fig. \ref{fig:DiscDelta}, the regions where $\mathcal{D}_\Delta\geq 0$, have been shown, for both cases of $\alpha\leq0$ and $\alpha>0$. 
As can be observed from the diagrams, the case of $\alpha \leq 0$ distinguishes between a black hole and a naked singularity through an extremal case where the two black hole horizons merge. On the other hand, for $\alpha > 0$, we never encounter a naked singularity, as the black hole can have three horizons for certain values of $\alpha$. By merging two of these horizons, the black hole still retains an event horizon and a Cauchy horizon, which merge when $\mathcal{D}_{\Delta} = 0$, leaving the black hole with only one horizon for all other corresponding values of $\alpha$. Note that the values of the spin parameter $a = a_{\rm{ext}}$, at which the horizons merge, can be obtained from the equation $\mathcal{D}_{\Delta} = 0$, which yields
\begin{eqnarray}
a_{\rm{ext}} &=& \frac{1}{\sqrt{6}}\Biggl\{
\Biggl\llbracket
2 \sqrt[3]{2} M^4 + 2 M^2 \Biggl(486 \sqrt[3]{2}\, \alpha  (1-\gamma)^2 \gamma -\Biggl[a_0-2 M^6+2430 \alpha  (1-\gamma)^2 \gamma  M^4+59049 \alpha ^2 (1-\gamma)^4 \gamma ^2 M^2\Biggr]^{1/3}\,\Biggr)\nonumber\\
&&+\Biggl(2a_0-4 M^6+4860 \alpha  (1-\gamma)^2 \gamma  M^4+118098 \alpha ^2 (1-\gamma)^4 \gamma ^2 M^2\,\Biggr)^{2/3}\nonumber\\
&&-84 \alpha  (1-\gamma) \gamma  \Biggl(a_0-2 M^6+2430 \alpha  (1-\gamma)^2 \gamma  M^4+59049 \alpha ^2 (1-\gamma)^4 \gamma ^2 M^2\Biggr)^{1/3}
\Biggr\rrbracket\nonumber\\
&&\times
\Biggl\llbracket(1-\gamma) \Biggr[a_0-2 M^6+2430 \alpha  (1-\gamma)^2 \gamma  M^4+59049 \alpha ^2 (1-\gamma)^4 \gamma ^2 M^2\,\Biggr]^{1/3}\Biggr\rrbracket^{-1}
\Biggr\}^{1/2},
    \label{eq:aext}
\end{eqnarray}
in which $a_0=9 M^2 \sqrt{3\alpha  (1-\gamma)^2 \gamma   \bigl[243 \alpha  (1-\gamma)^2 \gamma-4 M^2 \bigr]^3}$. For each fixed value of $\alpha$ and $\gamma$, if $a>\alpha_{\rm{ext}}$ holds, one can expect a naked singularity for $\alpha\leq 0$, and a one-horizon black hole for $\alpha>0$. The conditions stated above, can be observed in the behavior of the $\Delta$ function for fixed $\alpha$ and $\gamma$, as shown in Fig. \ref{fig:Delta}, where different values of the spin parameter have been taken into account.
As we can observe, for \( \alpha \leq 0 \), the equation \( \Delta = 0 \) has two real positive roots when \( a < a_{\rm{ext}} \), and one real positive root when \( a = a_{\rm{ext}} \). For \( a > a_{\rm{ext}} \), no real positive roots exist. In the case of \( \alpha > 0 \), as inferred from Fig. \ref{fig:DiscDelta}(b), there are two distinct values for \( a_{\rm{ext}} \), namely \( a_{\rm{ext}_1} \) and \( a_{\rm{ext}_2} \). When \( a_{\rm{ext}_1} < a < a_{\rm{ext}_2} \), there are three real positive roots for \( \Delta = 0 \), corresponding to a black hole with three horizons. At \( a = a_{\rm{ext}_1} \) or \( a = a_{\rm{ext}_2} \), two of these roots coalesce, leaving two positive roots. When \( a < a_{\rm{ext}_1} \) or \( a > a_{\rm{ext}_2} \), the two remaining roots also merge, resulting in a black hole with a single horizon.
It is also informative to show, graphically, the behavior of the equation $\Delta=0$ within both scenarios of the $\alpha$-parameter. In Fig. \ref{fig:Delta=0}, we have plotted the profile of this equation for different values of the $\alpha$-parameter, regarding changes in the spin parameter, when $\gamma$ is fixed. 
It can be directly inferred from the diagrams that, for $\alpha \leq 0$, the equation $\Delta = 0$ has two real positive roots, which merge into a single point as the spin parameter increases, corresponding to the EBH. In contrast, for $\alpha > 0$, the equation $\Delta = 0$ yields three real positive roots for small values of the spin parameter. As the spin parameter increases, the number of real positive roots decreases, eventually leaving only one horizon for larger values of the spin parameter.

One can also calculate analytically the roots of $\Delta=0$, which give
\begin{eqnarray}
    && r_1 = \frac{4}{1-\gamma} \sqrt{\frac{\chi_2}{3}}\,\cos\left(\frac{1}{3}\arccos\left(
    \frac{3\chi_3}{\chi_2}\sqrt{\frac{3}{\chi_2}}\,
    \right)-\frac{4\pi}{3}\right)+\frac{2M}{3(1-\gamma)},\label{eq:r1}\\
    && r_2 = \frac{4}{1-\gamma} \sqrt{\frac{\chi_2}{3}}\,\cos\left(\frac{1}{3}\arccos\left(
    \frac{3\chi_3}{\chi_2}\sqrt{\frac{3}{\chi_2}}\,
    \right)-\frac{2\pi}{3}\right)+\frac{2M}{3(1-\gamma)},\label{eq:r2}\\
    && r_3 = \frac{4}{1-\gamma} \sqrt{\frac{\chi_2}{3}}\,\cos\left(\frac{1}{3}\arccos\left(
    \frac{3\chi_3}{\chi_2}\sqrt{\frac{3}{\chi_2}}\,
    \right)\right)+\frac{2M}{3(1-\gamma)},\label{eq:r3}
\end{eqnarray}
where
\begin{subequations}
    \begin{eqnarray}
        && \chi_2 = -\frac{1}{12} \Bigl[3 a^2 (1-\gamma)+42 \alpha\gamma  (1-\gamma)  -4 M^2\Bigr],\\
        && \chi_3 = -\frac{M}{216} \Bigl[9 a^2 (1-\gamma)-9 \alpha  \gamma  \left(27 \gamma ^2-40 \gamma +13\right)-8 M^2\Bigr].
    \end{eqnarray}
    \label{eq:chi23}
\end{subequations}
Note that the correspondence of these roots to the black hole horizons depends strictly on the sign of the $\alpha$-parameter. For $\alpha \leq 0$, when all roots are real, the black hole possesses a Cauchy and an event horizon, which are determined respectively by $r_- = r_2$ and $r_+ = r_3$, with the other root being negative, i.e., $r_1 < 0$. As discussed before, the EBH occurs when $r_- = r_+$, beyond which a naked singularity arises. On the other hand, for $\alpha > 0$, there exists an intermediate horizon $r_- = r_1 < r_{\mathrm{in}} = r_2 < r_+ = r_3$ between the inner Cauchy horizon and the event horizon, when the conditions are favorable. This horizon disappears for $a = a_{\rm{ext}_{1,2}}$, and finally, $r_-$ also disappears when the above conditions are not satisfied, leaving a black hole with only one horizon $r_+$. 

\begin{figure}[h]
    \centering
    \includegraphics[width=5.3cm]{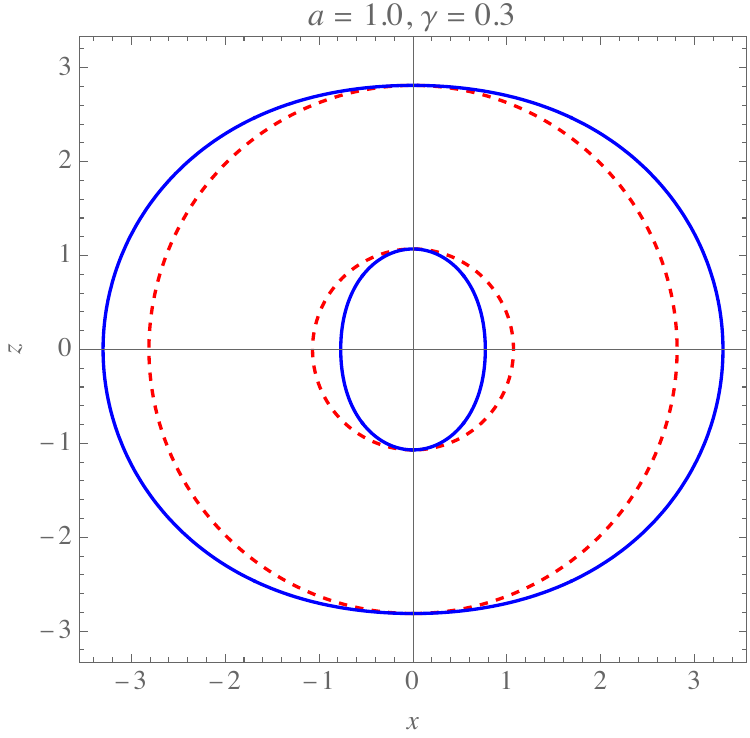} (a)\quad
    \includegraphics[width=5.3cm]{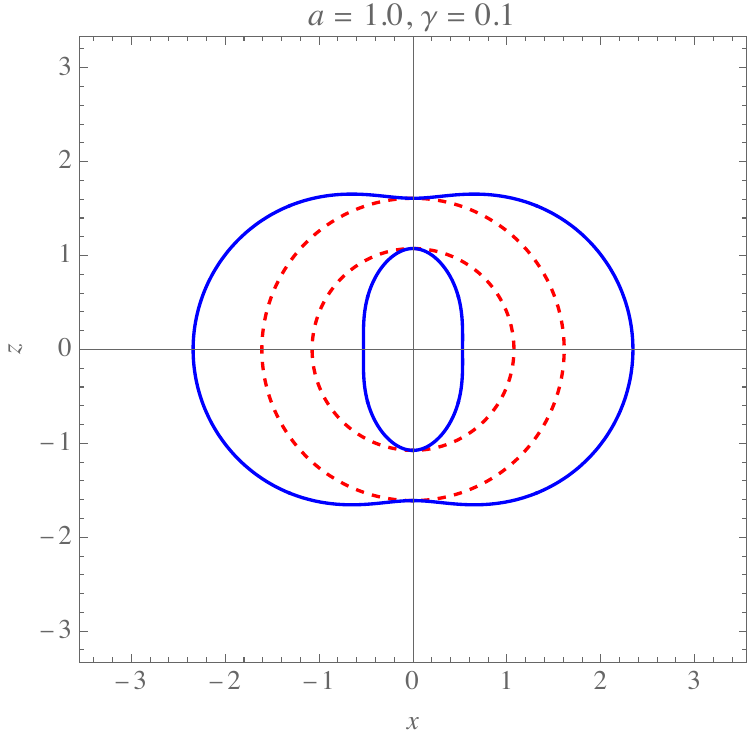} (b)\quad
    \includegraphics[width=5.3cm]{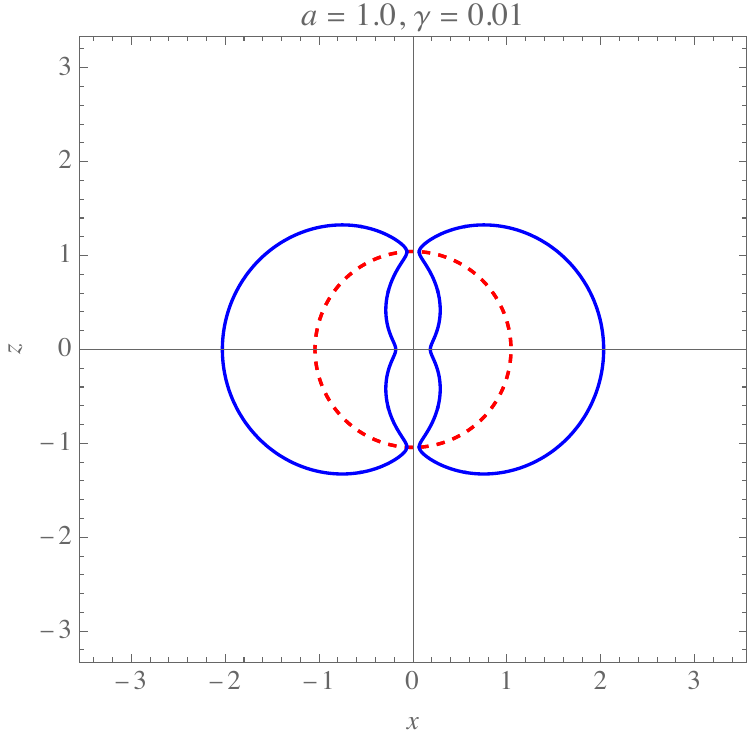} (c)
    \includegraphics[width=5.3cm]{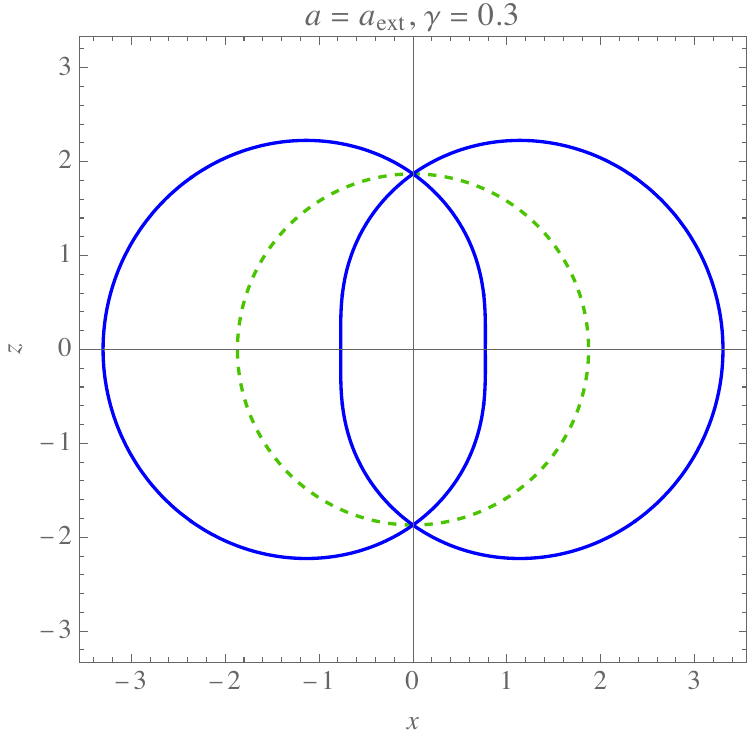} (d)\quad
    \includegraphics[width=5.3cm]{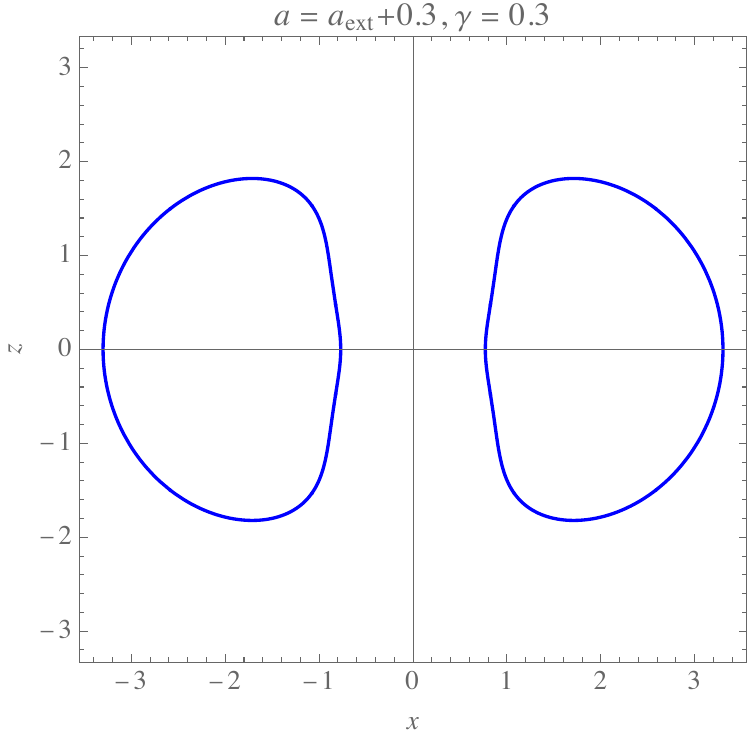} (e)
    \caption{The shapes of the ergosurfaces in the $z$--$x$ plane plotted for $\alpha = -0.4$ and different values of the spin parameter and the $\gamma$-parameter. The cases (a)--(c) correspond to the black hole with two horizons (the red dashed circles), the case (d) to the EBH with only one horizon (the green circle) for which $a_{\rm{ext}} = 1.348$, and the case (e) to a naked singularity. The unit of length along the axes is $M$.}
    \label{fig:ergo_n}
\end{figure}

\begin{figure}[h]
    \centering
    \includegraphics[width=5.3cm]{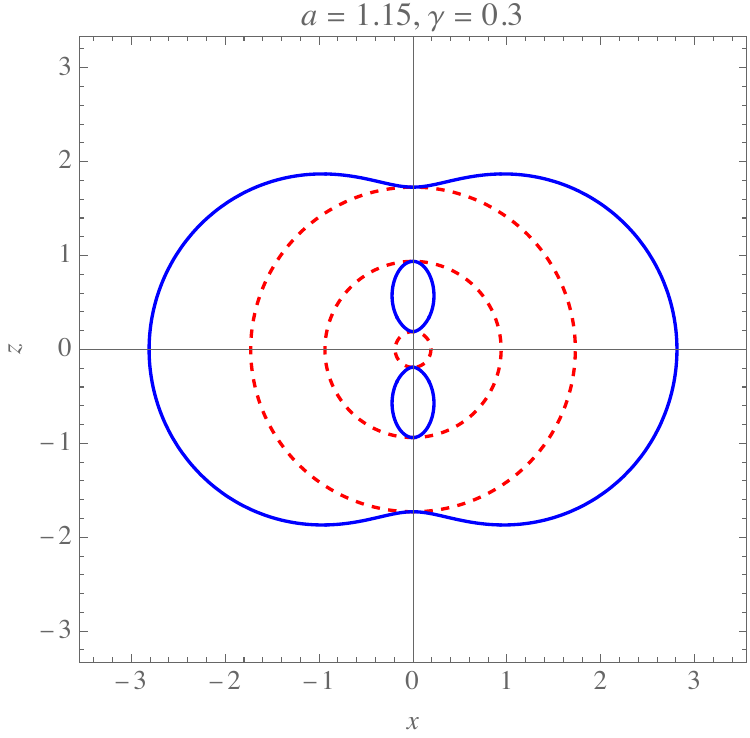} (a)\quad
    \includegraphics[width=5.3cm]{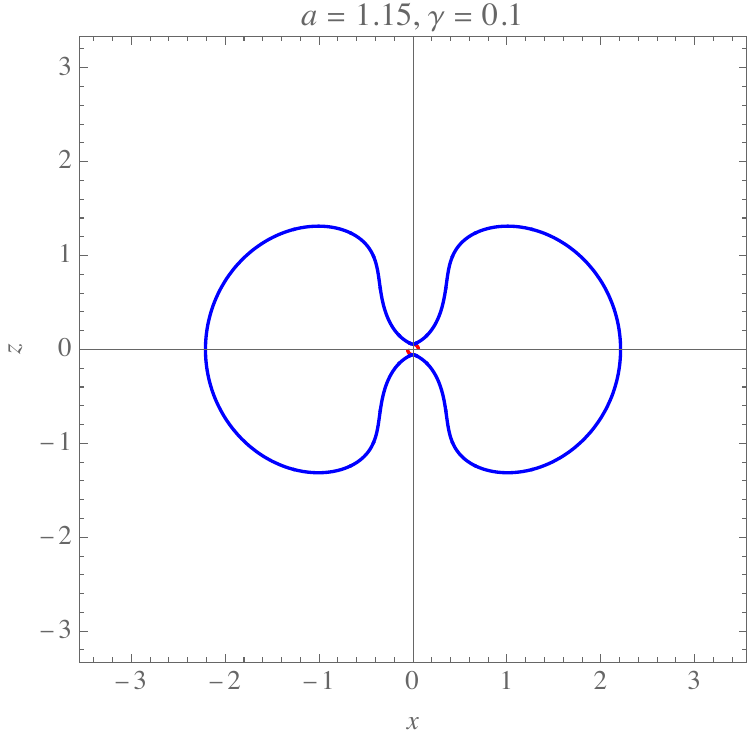} (b)\quad
    \includegraphics[width=5.3cm]{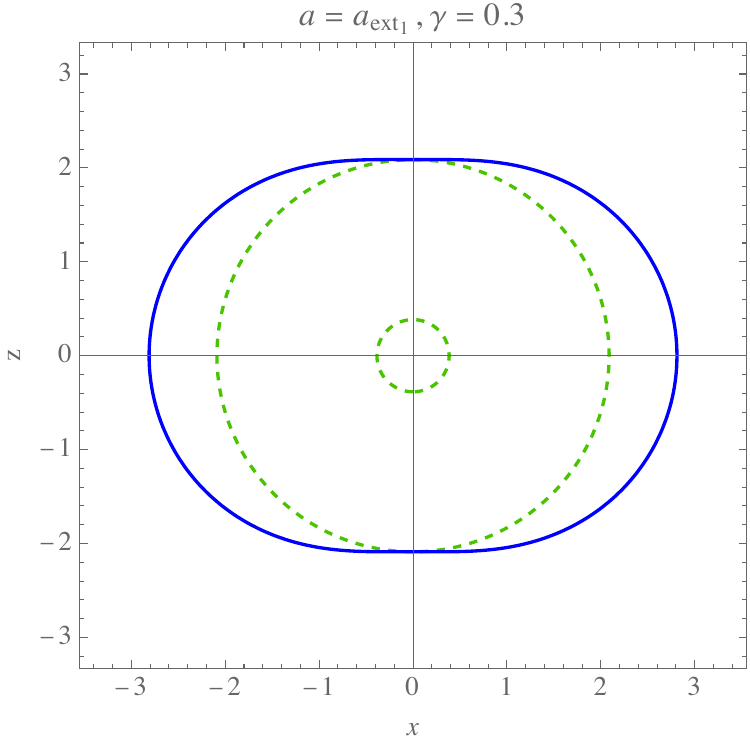} (c)
    \includegraphics[width=5.3cm]{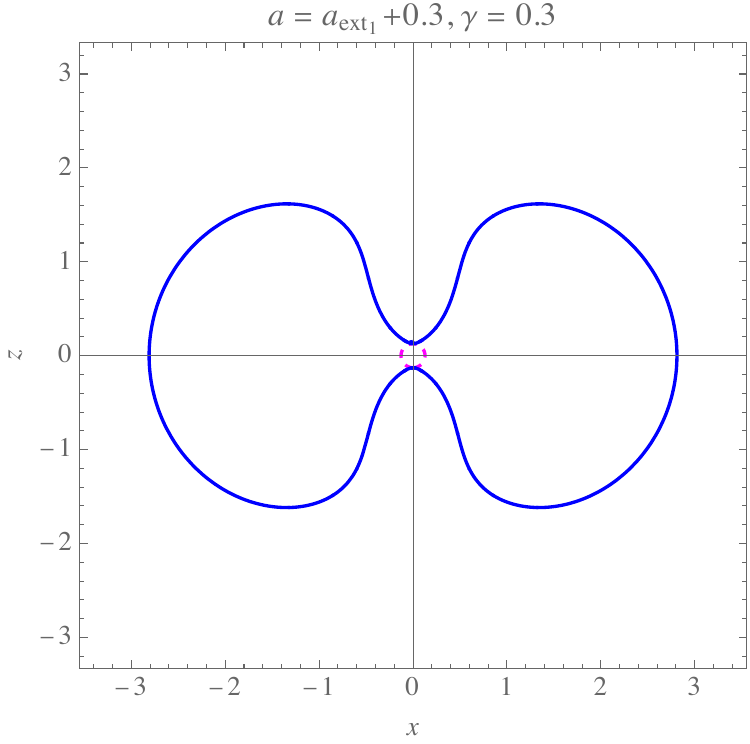} (d)\quad
    \includegraphics[width=5.3cm]{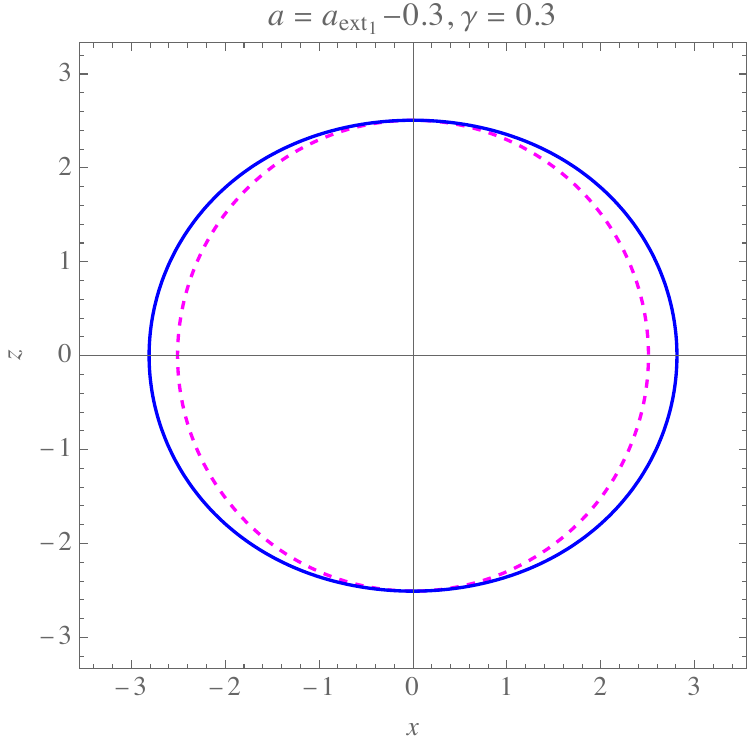} (e)
    \caption{The shapes of the ergosurfaces in the $z$--$x$ plane plotted for $\alpha = 0.04$ and different values of the spin parameter and the $\gamma$-parameter. The cases correspond to: (a) a black hole with three horizons (the dashed red curves), (b) a black hole with the same parameters as in case (a) but with $\gamma = 0.1$ (nearly naked singularity, similar to the KBH), (c) a black hole with two horizons (the green curves), in accordance with $a = a_{\rm{ext}_1} = 1.029$ in Fig. \ref{fig:Delta}(b), (d) a one-horizon black hole for $a = a_{\rm{ext}_1} + 0.3$ (again a nearly naked singularity, similar to the KBH), and (e) a one-horizon black hole for $a = a_{\rm{ext}_1} - 0.3$. The unit of length along the axes is $M$.}
    \label{fig:ergo_p}
\end{figure}

It is important to emphasize that, in addition to horizons, stationary black holes exhibit other hypersurfaces where static observers, whose world-lines are expressed as $\bm{u} = (-g_{tt})^{-1/2} \bm{\xi}^{t}$, cannot exist. Here, $\bm{\xi}^t$ represents the time-like Killing vector of the spacetime, aligned with the time coordinate. These hypersurfaces are known as static limits, where the observer's four-velocity $\bm{u}$ becomes null. The radii of these static limits, $r_{\rm{sl}}$, can be obtained by replacing $a \rightarrow a \cos\theta$ in the solutions \eqref{eq:r1}--\eqref{eq:r3}. In this sense, we obtain two hypersurfaces defined by the radii $r_{\rm{sl}_+}$ and $r_{\rm{sl}_-}$, which constrain the regions where a static observer can exist. These surfaces obey the hierarchy $r_{\rm{sl}_+} \geq r_+ \geq r_- \geq r_{\rm{sl}_-}$. Thus, the outer ergoregion is identified as $r_+ \leq r \leq r_{\rm{sl}_+}$, while the inner ergoregion is identified by $r_{\rm{sl}_-} \leq r \leq r_-$. Within these regions, no static observer can exist. Considering the Cartesian coordinates $x = r\sin\theta\cos\phi$, $y = r\sin\theta\sin\phi$, and $z = r\cos\theta$, Figs. \ref{fig:ergo_n} and \ref{fig:ergo_p} demonstrate several examples of the ergoregions for negative and positive values of the $\alpha$-parameter, respectively.
According to the diagrams, for $\alpha < 0$, the interior ergoregions become more similar to those of the KBH as we reduce the $\gamma$-parameter. However, the extremality for the RMBH shows significant differences from that of a KBH. The same holds for the corresponding naked singularity. When $\alpha > 0$, the significant differences are more apparent for the three-horizon black holes, as the interior ergosurface never forms inside the smallest horizon. Once again, by decreasing the value of $\gamma$, the spacetime mimics the exterior geometry of a KBH, which, in the examined cases, corresponds to a naked singularity. For the cases of $a < a_{\rm{ext}_1}$ and $a > a_{\rm{ext}_2}$, as demonstrated in Fig. \ref{fig:Delta}(b), a one-horizon black hole is formed, which is small for the former case and significantly larger for the latter case.

Now that the spacetime structure of the RMBH has been studied in detail, beginning from the next section, we turn our attention to the parametrization of null geodesics in the exterior geometry of the black hole, which has been obtained within the MNJA, through the method of separation of the Hamilton-Jacobi equation.

\section{Separation of the Hamilton-Jacobi equation and null geodesics}\label{sec:H_J}

It has been shown in Ref. \cite{Shaikh:2020} that the stationary spacetimes obtained from the NJA, in the case that they can be presented in Boyer-Lindquist-like coordinates, will also admit the separability of the Hamilton-Jacobi equation. For the case of $A(r) = B(r)$, this separability has also been studied in Ref. \cite{Azreg:2014} in the context of the MNJA, and for the case of $A(r) \neq B(r)$, this separability has been studied in Ref. \cite{junior_spinning_2020}. There, the authors argue that although, due to the complexity of \eqref{eq:Psi_cond_1}, it is a formidable task to determine a general solution for $\Psi$ when $A(r) \neq B(r)$, this function, however, does not need to be identified explicitly in the study of null geodesics, since the separability of the Hamilton-Jacobi equation is independent of this function. In fact, the Hamilton-Jacobi equation is given by
\begin{equation}
   \frac{\partial\mathcal{S}}{\partial\tau}+\mathcal{H}=0,
   \label{eq:HJ_0}
\end{equation}
in which $\mathcal{S}$ is the Jacobi action, $\tau$ is the affine parameter of the geodesic curves, and $\mathcal{H}$ is the canonical Hamiltonian, expressed as
\begin{equation}
\mathcal{H}=\frac{1}{2} g^{\mu\nu} p_\mu p_\nu,
    \label{eq:HJ_1}
\end{equation}
with $\bm{p}$ being the conjugate momentum covector, which is defined as
\begin{equation}
p_\mu=\frac{\partial\mathcal{S}}{\partial x^{\mu}}.
    \label{eq:HJ_2}
\end{equation}
In this manner, the constants of motion can be defined by exploiting the spacetime symmetries associated with the \( t \) and \( \phi \)-coordinates. These constants are given by \( E = -p_t \) and \( L = p_\phi \), where \( E \) represents the energy and \( L \) denotes the angular momentum of the test particles. Consequently, we can adopt the ansatz
\begin{equation}
\mathcal{S}=\frac{m^2}{2}\tau-Et+L\phi+\mathcal{S}_r(r)+\mathcal{S}_\theta(\theta),
    \label{eq:HJ_3}
\end{equation}
for the Jacobi action, in which $m$ is the test particles' mass. By applying this ansatz, along with the definitions given in \eqref{eq:HJ_1} and \eqref{eq:HJ_2}, and the line element in \eqref{eq:ds_kerr_like}, the Hamilton-Jacobi equation \eqref{eq:HJ_0} yields
\begin{equation}
-\frac{1}{\Psi\Delta}\left[E\left(K+a^2\right)-aL\right]^2+\frac{p_\theta^2}{\Psi}+\frac{1}{\Psi\sin^2\theta}\left(aE\sin^2\theta-L\right)^2+\frac{\Delta\, p_r^2}{\Psi}=-m^2.
     \label{eq:HJ_4}
\end{equation}
For null geodesics, it is naturally \( m = 0 \), and hence the conformal factor \( \Psi \) does not contribute. Accordingly, we can infer the equation
\begin{equation}
\Delta\, p_r^2 - \frac{1}{\Delta}\left[E\left(K + a^2\right) - aL\right]^2 = -\left[p_\theta^2 + \frac{1}{\sin^2\theta}\left(aE\sin^2\theta - L\right)^2\right] \equiv \mathcal{K},
\label{eq:HJ-5}
\end{equation}
where we have separated the \( r \)-dependent segment of the equation from the \( \theta \)-dependent part, using a constant \( \mathcal{K} \) defined as
\begin{equation}
\mathcal{K} = \Q + \left(aE - L\right)^2,
\label{eq:mK}
\end{equation}
with \( \Q \) representing Carter's constant \cite{Carter:1968}. Based on this method, the equations for \( p_r \) and \( p_\theta \) are obtained as
\begin{eqnarray}
p_r^2 &=& \frac{1}{\Delta^2} \left[E\left(K + a^2\right) - a L\right]^2 - \frac{1}{\Delta}\left[\Q + \left(aE - L\right)^2\right], \label{eq:pr} \\
p_\theta^2 &=& \Q + a^2 E^2 \cos^2\theta - L^2 \cot^2\theta. \label{eq:ptheta}
\end{eqnarray}
As we can see from the above equations, the Hamilton-Jacobi equation is fully separable for null geodesics in the line element \eqref{eq:ds_kerr_like}. It is important to highlight that the NJA admits the separability of the Hamilton-Jacobi equations only if the rotating spacetime's line element can be written in Boyer-Lindquist-like coordinates. In the context of the MNJA, however, this separability always holds for null geodesics. Now, since \( p_\mu = g_{\mu\nu} u^\nu = g_{\mu\nu}\left({d x^\nu}/{d\tau}\right) \), with the help of Eqs. \eqref{eq:pr} and \eqref{eq:ptheta}, we can write the full set of equations of motion for null geodesics as follows:
\begin{eqnarray}
\Psi \frac{d t}{d\tau} &=& \frac{E}{\Delta} \left[ \left(K + a^2\right)^2 - a^2 \Delta \sin^2\theta - a\left(K + a^2 - \Delta\right) \xi \right], \label{eq:tdot} \\
\Psi \frac{d r}{d\tau} &=& E \sqrt{\mathcal{R}(r)}, \label{eq:rdot} \\
\Psi \frac{d \theta}{d\tau} &=& E \sqrt{\Theta(\theta)}, \label{eq:thetadot} \\
\Psi \frac{d\phi}{d\tau} &=& \frac{E}{\Delta \sin^2\theta} \left[ \left(\Delta - a^2 \sin^2\theta\right) \xi + a \sin^2\theta \left(K + a^2 - \Delta\right) \right], \label{eq:phidot}
\end{eqnarray}
where
\begin{subequations}
\begin{eqnarray}
\xi &=& \frac{L}{E}, \label{eq:xi_def} \\
\eta &=& \frac{\Q}{E^2}, \label{eq:eta_def}
\end{eqnarray}
\end{subequations}
are the gauge-invariant constants of motion, and we have also defined
\begin{subequations}
\begin{eqnarray}
\mathcal{R}(r) &=& \left(K + a^2 - a\xi\right)^2 - \Delta \left[\eta + \left(a - \xi\right)^2\right], \label{eq:mR} \\
\Theta(\theta) &=& \eta + \cos^2\theta \left(a^2 - \frac{\xi^2}{\sin^2\theta}\right). \label{eq:Theta}
\end{eqnarray}
\end{subequations}
It is essential to rely on the conditions \( R(r) \geq 0 \) and \( \Theta(\theta) \geq 0 \) for the null trajectories to exist in the exterior geometry of the black hole.

\subsection{Orbits of constant radius and the photon regions}\label{subsec:photonregions}

In curved spacetimes, understanding particle and photon trajectories involves analyzing their radial motion. For light rays, this analysis centers on the radial equation of motion, leading to different types of orbits, such as deflecting, captured, and spherical (critical) orbits. Spherical photon orbits, in particular, represent paths where the radial coordinate remains constant, skimming the black hole’s horizon. These orbits are critical because they determine whether a photon escapes or plunges into the black hole, and they are inherently unstable, forming an infinite sequence of photon rings that define the black hole's shadow. In static spacetimes like Schwarzschild, these orbits are planar; however, in rotating black holes, frame-dragging creates a photon region where spherical orbits become non-planar. This region, bounded by the innermost and outermost circular orbits, defines the photon orbits, as extensively studied in KBH solutions (see Refs. \cite{Chandrasekhar:2002,Bardeen:1972a,Bardeen:1973b}). Research on these orbits and their observational significance has expanded significantly in Kerr and Kerr-like spacetimes (see Refs. \cite{stoghianidis_polar_1987,cramer_using_1997,Teo:2003,Johannsen:2013,Grenzebach:2014,Perlick:2017,charbulak_spherical_2018,Johnson_universal_2020,Himwich:2020,Gelles:2021,Ayzenberg:2022,Das:2022,fathi_spherical_2023,ANJUM2023101195,Chen:2023,andaru_spherical_2023}).

\begin{figure}[h]
    \centering
    \includegraphics[width=5.3cm]{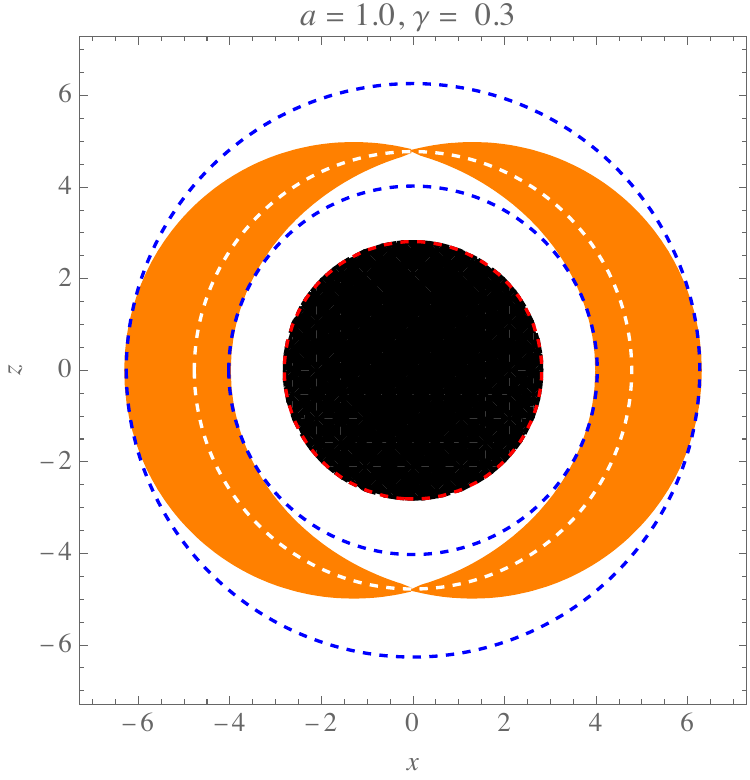} (a)\quad
   \includegraphics[width=5.3cm]{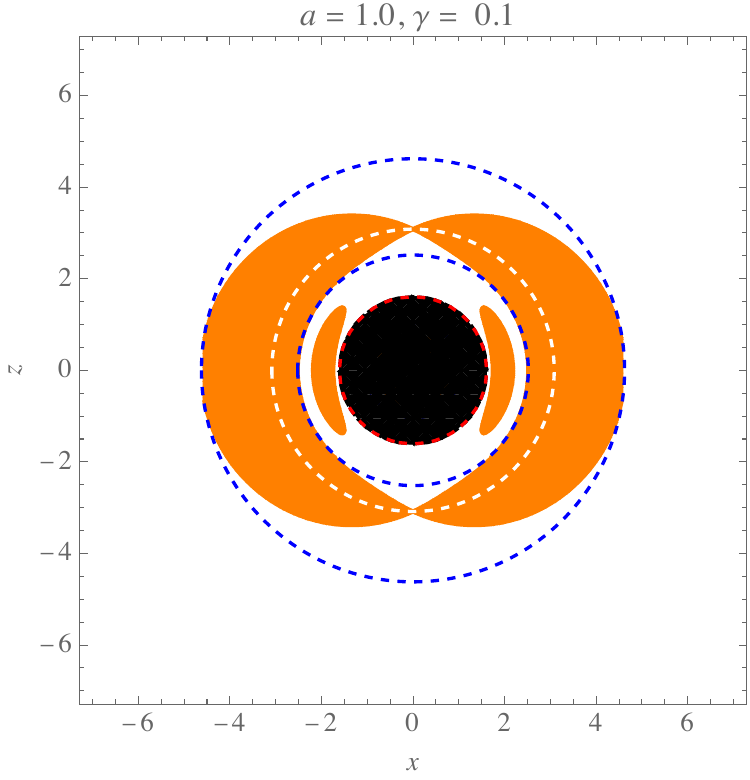} (b)\quad
    \includegraphics[width=5.3cm]{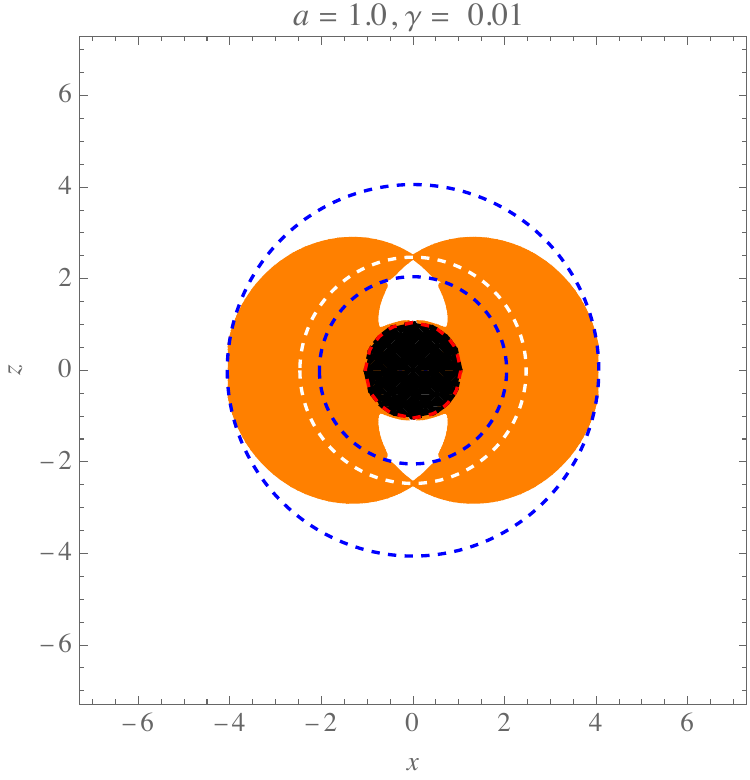} (c)
    \includegraphics[width=5.3cm]{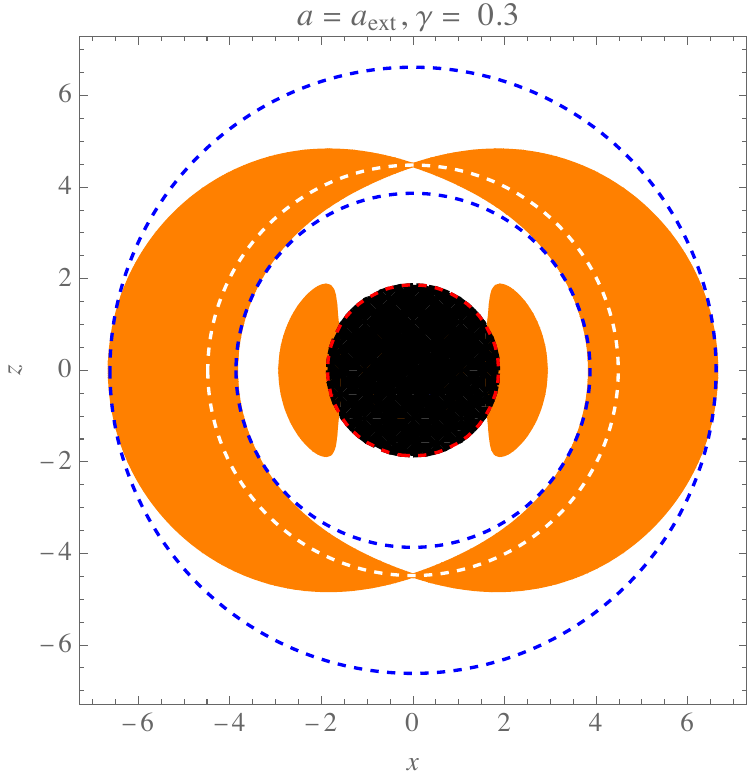} (d)\quad
    \includegraphics[width=5.3cm]{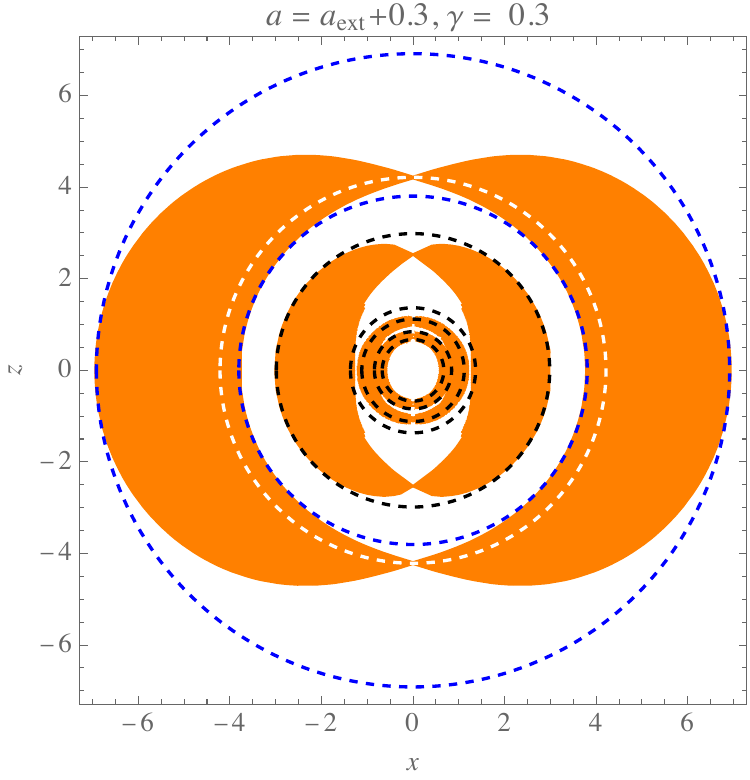} (e)
    \caption{The photon regions around the RMBH in the $z$--$x$ plane (indicated by orange areas) plotted for $\alpha = -0.4$ and various values of the spin parameter and the $\gamma$-parameter, consistent with the configurations shown in Fig. \ref{fig:ergo_n}. The blue dashed circles represent the radii of exterior planar photon orbits, denoted as $r_{p_\mp}$. The white dashed circles correspond to a radius of $r_0$, while the central black disk has a radius of $r_+$. The panels display the following values: (a) $r_+ = 2.812$, $r_{p_-} = 4.024$, $r_{p_+} = 6.263$, and $r_0 = 4.777$; (b) $r_+ = 1.610$, $r_{p_-} = 2.518$, $r_{p_+} = 4.620$, and $r_0 = 3.083$; (c) $r_+ = 1.043$, $r_{p_-} = 2.044$, $r_{p_+} = 4.057$, and $r_0 = 2.471$; (d) $r_+ = 1.870$, $r_{p_-} = 3.866$, $r_{p_+} = 6.620$, and $r_0 = 4.483$. Panel (e) illustrates the photon regions around a naked singularity, bounded by blue curves corresponding to $r_{p_-} = 3.804$ and $r_{p_+} = 6.915$, which are comparable to the exterior photon regions for black holes. In this case, smaller photon regions are enclosed by black dashed circles with radii (from largest to smallest) $r_{p_1} = 2.985$, $r_{p_2} = 1.366$, $r_{p_3} = 1.118$, $r_{p_4} = 0.843$, and $r_{p_5} = 0.672$. For this example, $r_0 = 4.215$. The units along the axes are given in terms of $M$.}
    \label{fig:photon_region_n}
\end{figure}

\begin{figure}[h]
    \centering
    \includegraphics[width=5.3cm]{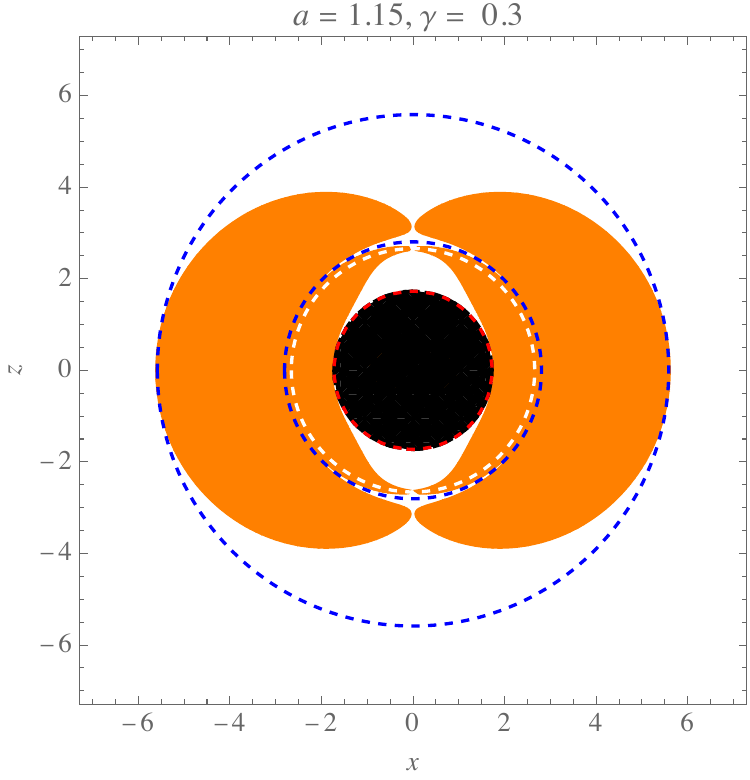} (a)\quad
   \includegraphics[width=5.3cm]{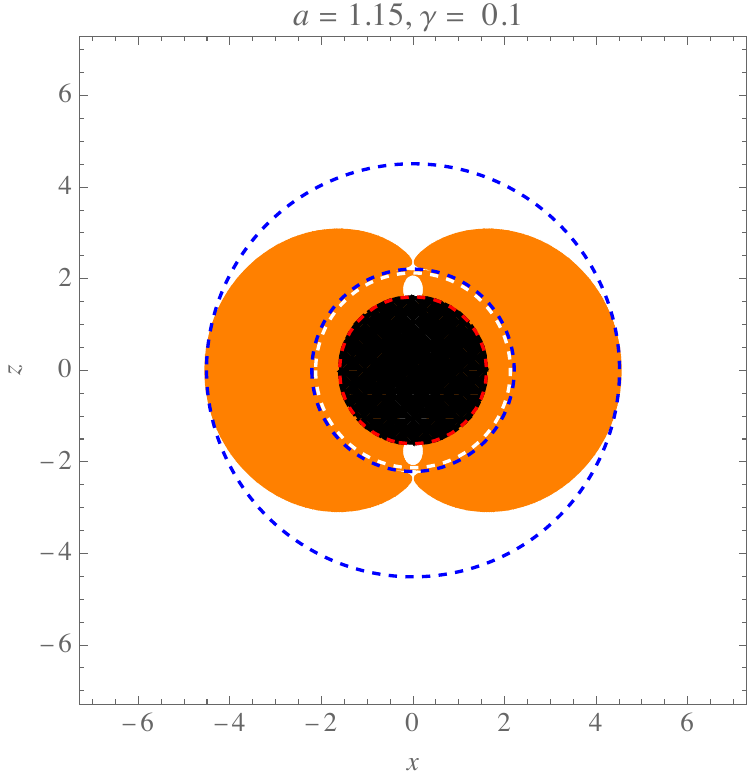} (b)\quad
    \includegraphics[width=5.3cm]{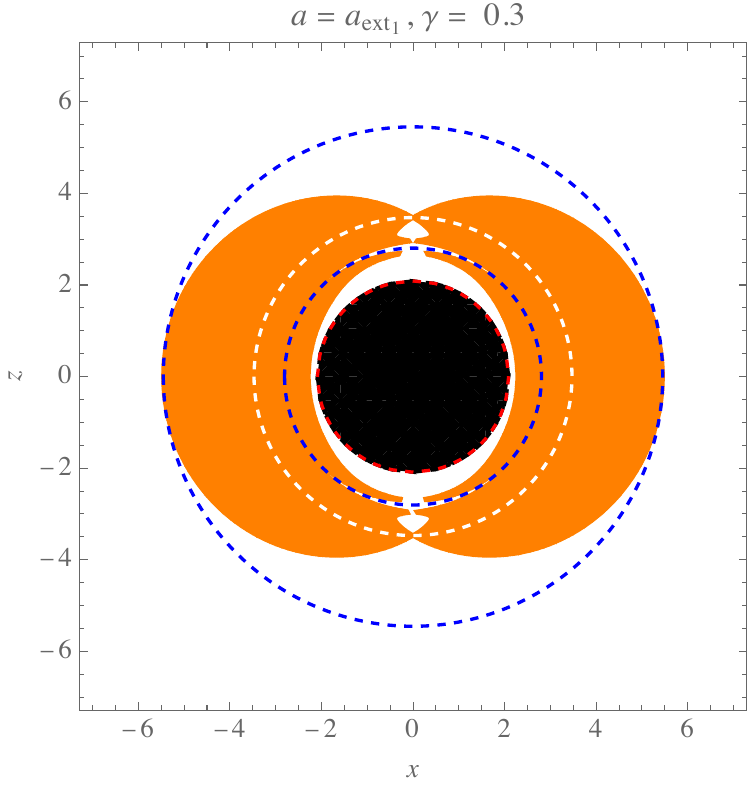} (c)
    \includegraphics[width=5.3cm]{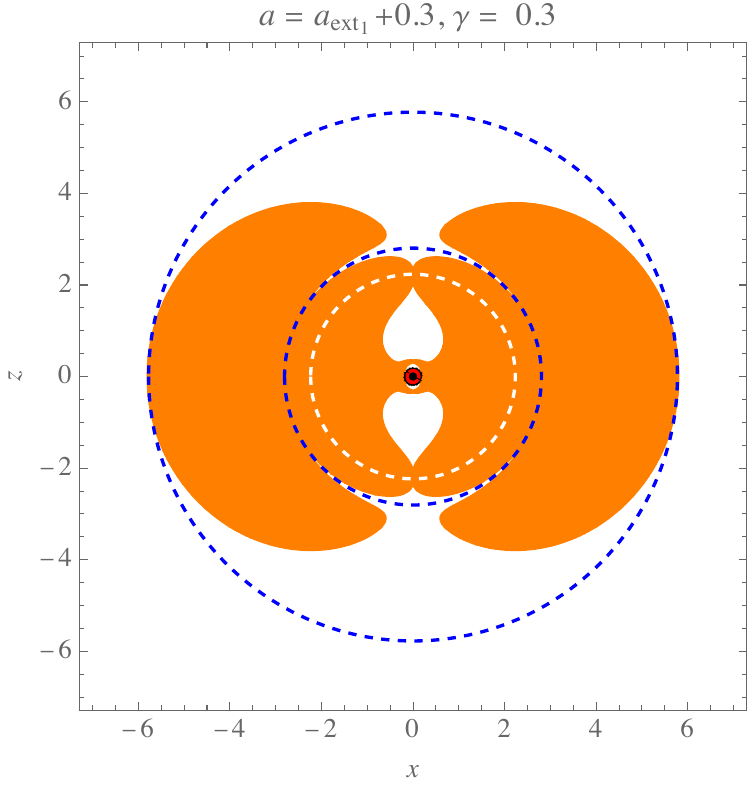} (d)\quad
    \includegraphics[width=5.3cm]{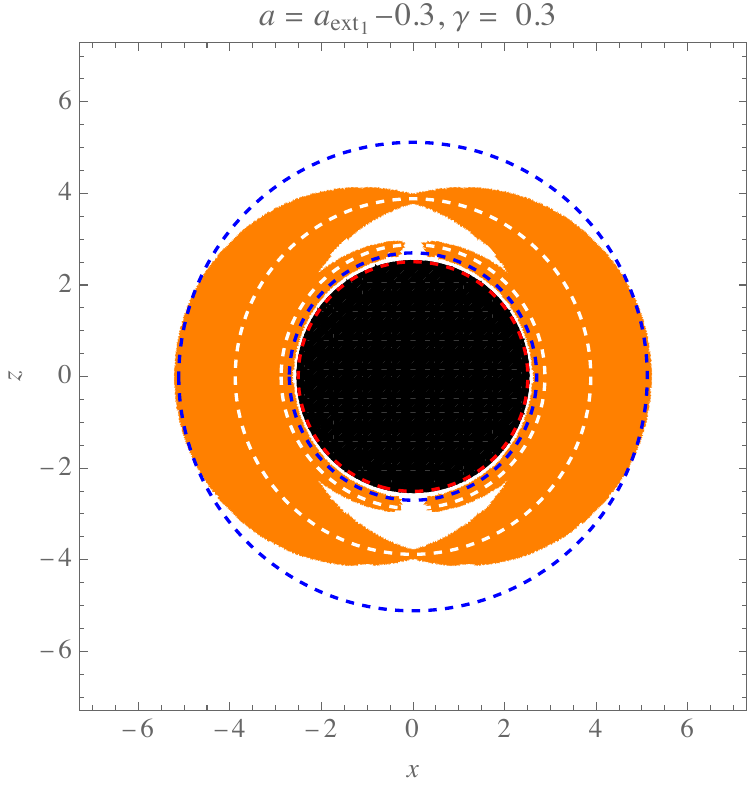} (e)
    \caption{The shapes of the photon regions around the RMBH in the $z$--$x$ plane (depicted in orange) plotted for $\alpha = 0.04$ and various values of the spin and $\gamma$-parameters, following the configurations provided in Fig. \ref{fig:ergo_p}. Only one of the values for $r_0$ has been considered and is displayed. The panels display the following values: (a) $r_+ = 1.729$, $r_{p_-} = 2.810$, $r_{p_+} = 5.849$, and $r_0 = 2.658$; (b) $r_+ = 1.610$, $r_{p_-} = 2.210$, $r_{p_+} = 4.511$, and $r_0 = 2.127$; (c) $r_+ = 2.100$, $r_{p_-} = 2.810$, $r_{p_+} = 5.454$, and $r_0 = 3.742$; (d) $r_+ = 0.128$, $r_{p_-} = 2.810$, $r_{p_+} = 5.775$, and $r_0 = 2.234$; and (e) $r_+ = 2.510$, $r_{p_-} = 3.883$, $r_{p_+} = 5.116$, and $r_0 = 2.881$. The unit of length along the axes is expressed in terms of $M$.}
    \label{fig:photon_region_p}
\end{figure}

In fact, spherical photon orbits at a given radius \( r_p \) are determined under the criteria \( \mathcal{R}(r_p) = 0 = \mathcal{R}'(r_p) \) \cite{Teo:2003}. Exploiting the form given in Eq. \eqref{eq:mR}, these equations provide the sets of solutions
\begin{eqnarray}
    && \xi_{p} = \left(\frac{K+a^2}{a}\right)_{r_p},\label{eq:xip_1}\\
    && \eta_{p} = -\left(\frac{K^2}{a^2}\right)_{r_p},\label{eq:etap_1}
\end{eqnarray}
and 
\begin{eqnarray}
    && \xi_{p} = \left(\frac{K+a^2}{a}-\frac{2 K' \Delta}{a \Delta'}\right)_{r_p},\label{eq:xip_2}\\
    && \eta_{p} = \left(\frac{4\Delta (K')^2}{(\Delta')^2}-\frac{1}{a^2}\left[K-\frac{2 K' \Delta}{\Delta'}\right]^2\right)_{r_p}.\label{eq:etap_2}
\end{eqnarray}
These equations determine the critical impact parameter for light rays on constant-\(r\) orbits. However, it is clear from Eq. \eqref{eq:etap_1} that \( \eta_p < 0 \), which leads to a non-physical result. Thus, we disregard the solutions in \eqref{eq:xip_1} and \eqref{eq:etap_1} and instead consider those in Eqs. \eqref{eq:xip_2} and \eqref{eq:etap_2} as the critical constants of motion for spherical photon orbits. Notably, for planar orbits with \( \theta = \pi/2 \), one can infer from Eqs. \eqref{eq:thetadot} and \eqref{eq:Theta} that \( \eta_p = 0 \). This result allows us to use Eq. \eqref{eq:etap_2} to obtain the radii of planar circular photon orbits.  For the RMBH spacetime, given the complexity of the metric functions \eqref{eq:A(r)} and \eqref{eq:B(r)}, this equation results in a polynomial of fifteenth order, rendering the radii of planar photon orbits analytically intractable. However, numerical solutions can be used to determine the radii of prograde and retrograde planar circular orbits, \( r_{p_-} \) and \( r_{p_+} \), respectively, which define the inner and outer boundaries of the photon region. The photon region is characterized by the condition \( \Theta(\theta) \geq 0 \) for spherical photon orbits, using the critical impact parameters in Eqs. \eqref{eq:xip_2} and \eqref{eq:etap_2} in the angular potential \eqref{eq:Theta}. By solving for the radii with \( \eta_p = 0 \), we illustrate examples of the photon regions forming around the RMBH, as shown in Figs. \ref{fig:photon_region_n} and \ref{fig:photon_region_p}, for various parameter values as exemplified in Figs. \ref{fig:ergo_n} and \ref{fig:ergo_p}.
In these diagrams, in addition to the radii of planar orbits $r_0$, we have also shown the radii of polar orbits, which indicate the spherical orbits for photons with no associated angular momentum. These photons traverse the entire polar angle, passing through the poles without any change in the azimuthal angle. The radii of such orbits are thus obtained by solving the equation $\xi_p=0$, using the expression in Eq. \eqref{eq:xip_2}. As observed from the diagrams, photon regions can form within $r_{p_-}$ for both cases, as these regions are bounded by the black hole's event horizon. Notably, photon regions can also form inside the event horizon (see, for example, Ref. \cite{Grenzebach:2014}), with these regions bounded by the radii of circular orbits within the black hole. When a naked singularity forms for $\alpha < 0$, these regions replicate, generating photon regions of decreasing size as they approach the singularity. For black holes with $\alpha > 0$, interesting features emerge, as the interior photon regions (those within $r_{p_-}$) vary in size and can form distinct branches, each containing a separate spherical polar orbit.

It is important to note that, for a distant observer, the photons residing within the photon regions (i.e., photons on spherical orbits) form the innermost photon rings, as they complete numerous half-orbits around the black hole before reaching the observer (for further discussion, see Refs. \cite{Gralla:2019,bisnovatyi-kogan_analytical_2022,tsupko_shape_2022}, as well as the foundational works in Refs. \cite{claudel_geometry_2001,virbhadra_relativistic_2009,virbhadra_compactness_2024}). These photons thus play a key role in defining the true boundary of the shadow (or the critical curve), which will be explored in the next subsection.

\subsection{The black hole shadow}\label{subsec:shadow}

The study of light propagation around black holes holds significant importance in astrophysics, especially given the insights enabled by advances in observational astronomy. Photons on unstable orbits in the gravitational field of a black hole either fall onto the event horizon or escape to infinity. To a distant observer, those that escape appear as a bright photon ring that bounds the black hole shadow \cite{Synge:1966,Cunningham:1972,Bardeen:1973a,Luminet:1979}. In particular, Luminet's 1979 optical simulation of a SBH and its accretion \cite{Luminet:1979} provided detailed insights into photon rings formed in the highly warped regions around black holes. These insights helped scientists refine the shadow of rotating black holes within their respective photon rings. Subsequently, mathematical techniques to determine the size and shape of a KBH's shadow were developed by Bardeen \cite{Bardeen:1972a,Bardeen:1973a,Bardeen:1973b}, later expanded upon by Chandrasekhar \cite{Chandrasekhar:579245}, and further generalized (see Refs.~\cite{Bray:1986,Vazquez:2004,Grenzebach:2014,Grenzebach:2016,Perlick:2018a,Kogan:2018}). These methods enabled rigorous analytical, numerical, and observational studies across various black hole spacetimes in general relativity and alternative theories of gravity \cite{Vries:1999,Kramer:2004,Shen:2005,Psaltis:2008,Harko:2009,Amarilla:2010,Amarilla:2012,Yumoto:2012,Amarilla:2013,Atamurotov:2013,Abdujabbarov:2015,Psaltis:2015,Abdujabbarov:2016,Johannsen:2016a,Amir:2018,Tsukamoto:2018,Cunha:2018,Mizuno:2018,Mishra:2019,Shaikh:2019,Psaltis:2019,Dymnikova:2019,Kumar:2020a,Kumar:2020b,junior_spinning_2020,fathi_ergosphere_2021,narzilloev_optical_2022,zahid_shadow_2023,nozari_asymptotically_2023-1,fathi_black_2024,zahid_shadow_2024,raza_shadow_2024,sanchez_shadow_2024,khan_optical_2024,jafarzade_kerrnewman_2024}. As a result, black hole shadows have become essential in understanding light propagation near event horizons. Recent studies have also explored correlations between black hole shadow properties and fundamental parameters, including thermodynamic quantities \cite{Zhang:2020,Belhaj:2020}.

Historically, various definitions have been proposed in the literature to describe the visual characteristics of black holes. Synge introduced the concept of the \textit{escape cone} \cite{Synge:1966}, while Zeldovich and Novikov referred to the \textit{cone of gravitational radiation capture} \cite{zeldovich_relativistic_1966}. Bardeen, Chandrasekhar, and Luminet further popularized terms like the \textit{optical appearance of black holes} and \textit{black hole image}, which are now widely used to describe how black holes appear visually \cite{Bardeen:1973b,Chandrasekhar:1998,Luminet:1979nyg,luminet_seeing_2018}. The term \textit{black hole shadow}, introduced by Falcke, Melia, and Agol \cite{Falcke_2000}, specifically refers to the dark region within the apparent boundary of the black hole, which is sometimes termed the \textit{photon ring} \cite{johannsen_testing_2010,Johnson_universal_2020} or the \textit{critical curve} \cite{gralla_measuring_2020} (see Ref.~\cite{perlick_calculating_2022} for further historical insights). In the context of the current study, the shadow is supposed to be bounded by the innermost light rings that represent lensed images of the luminous background.

\begin{figure}[t]
    \centering
    \includegraphics[width=5.3cm]{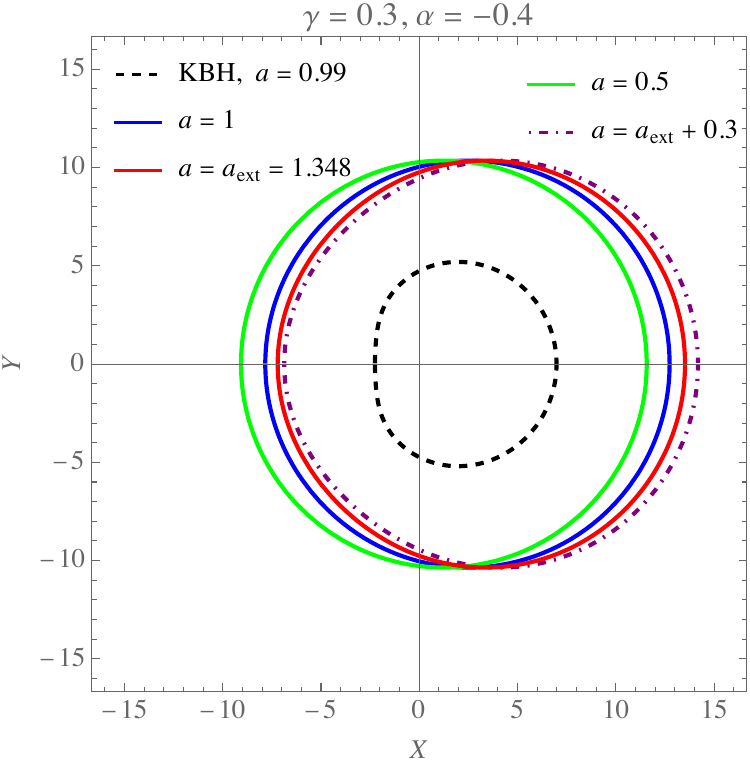} (a)\quad
   \includegraphics[width=5.3cm]{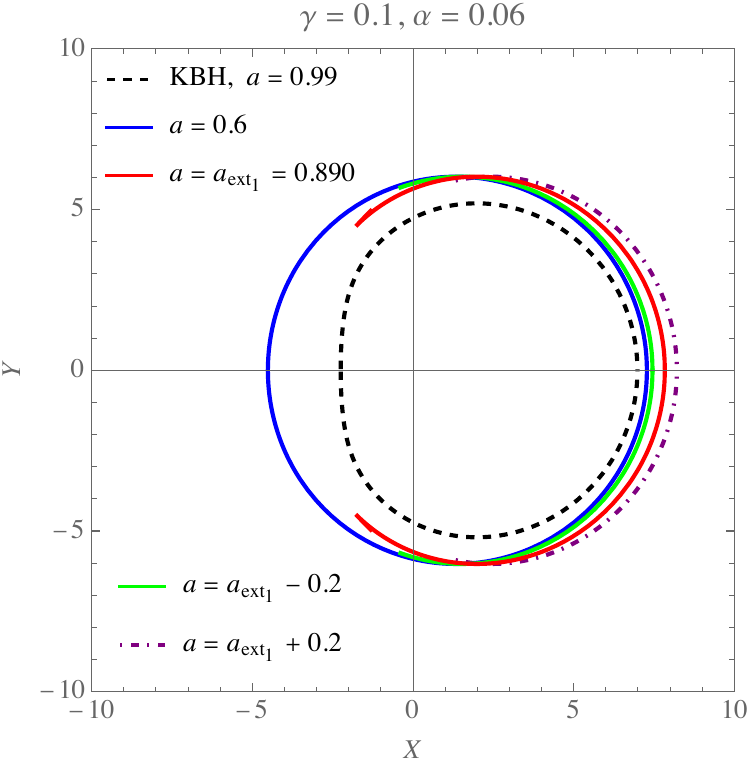} (b)\quad
    \includegraphics[width=5.3cm]{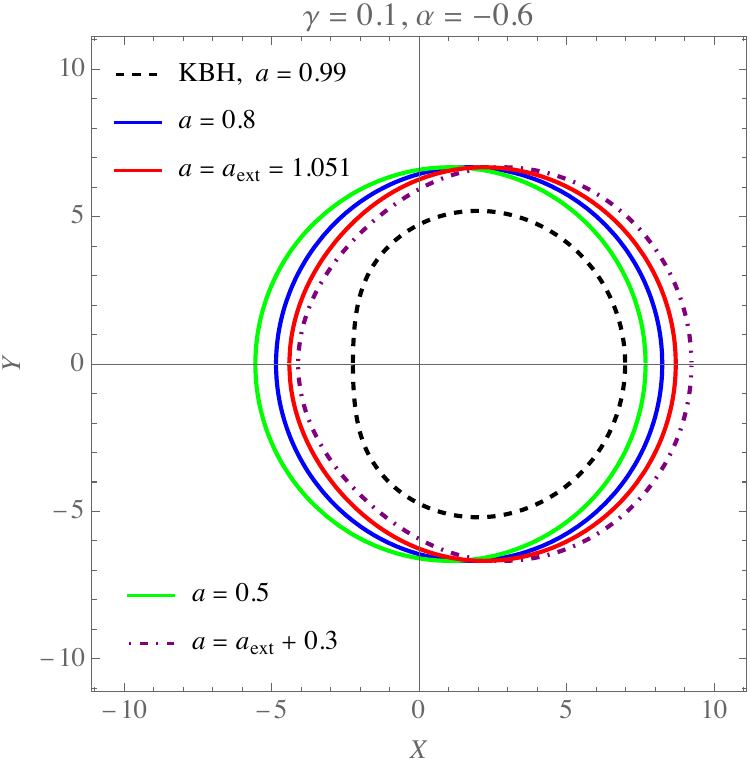} (c)
    \includegraphics[width=5.3cm]{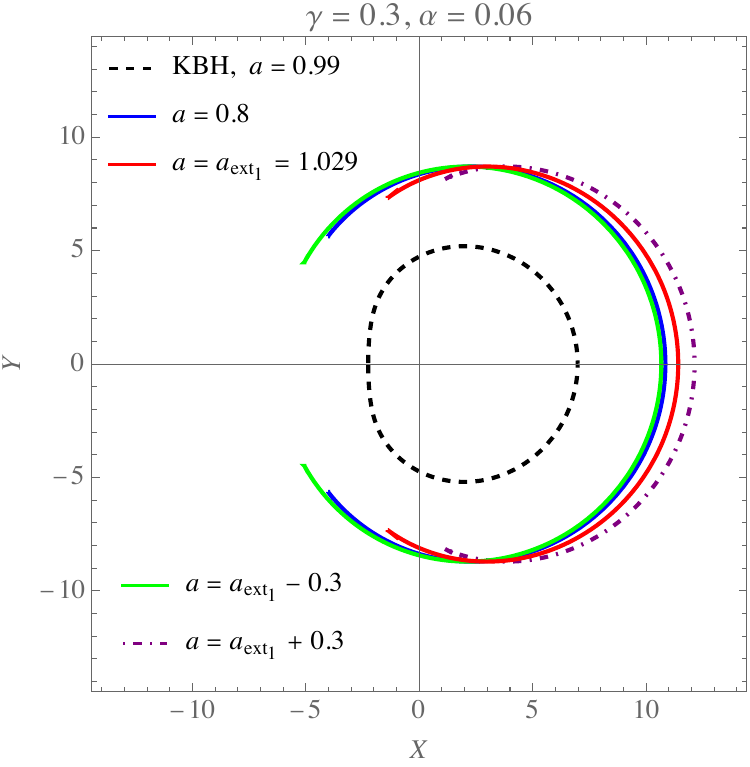} (d)\quad
    \includegraphics[width=5.3cm]{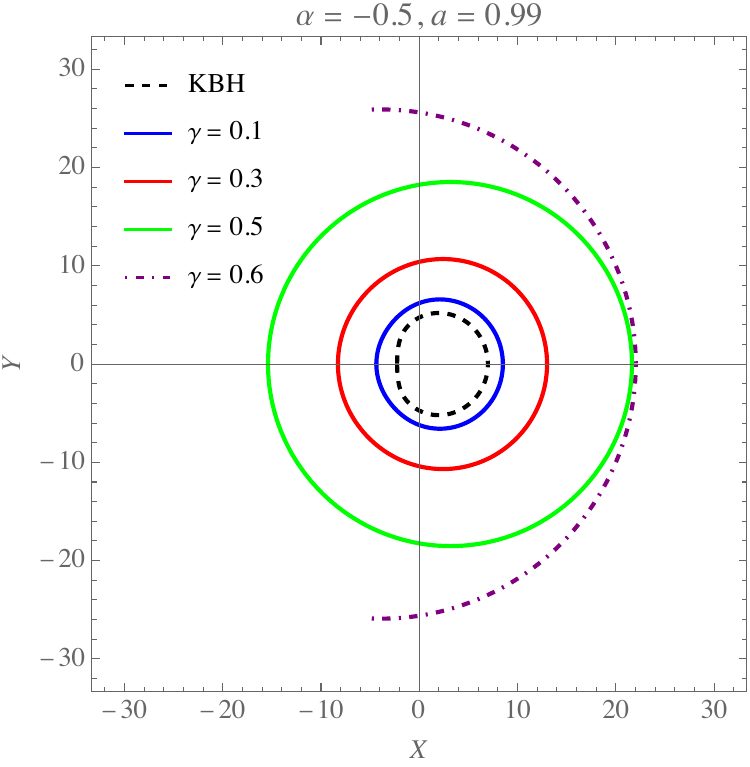} (e)\quad
    \includegraphics[width=5.3cm]{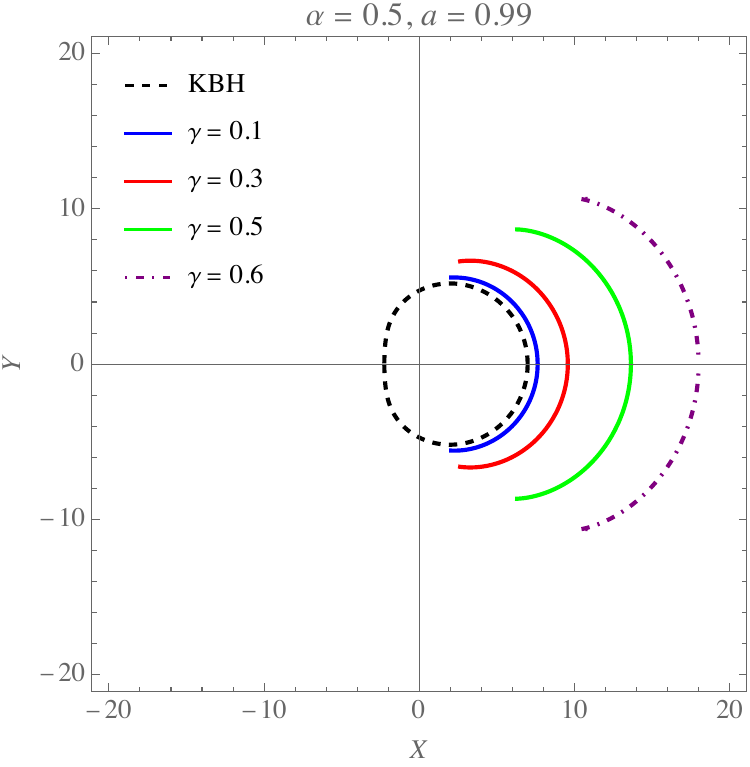} (f)
    \includegraphics[width=5.3cm]{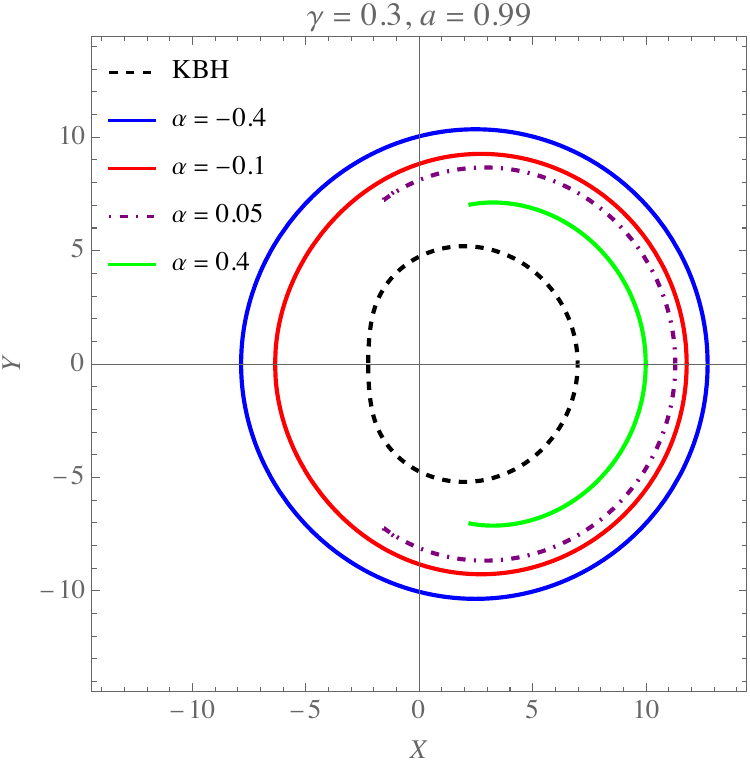} (g)\quad
    \includegraphics[width=5.3cm]{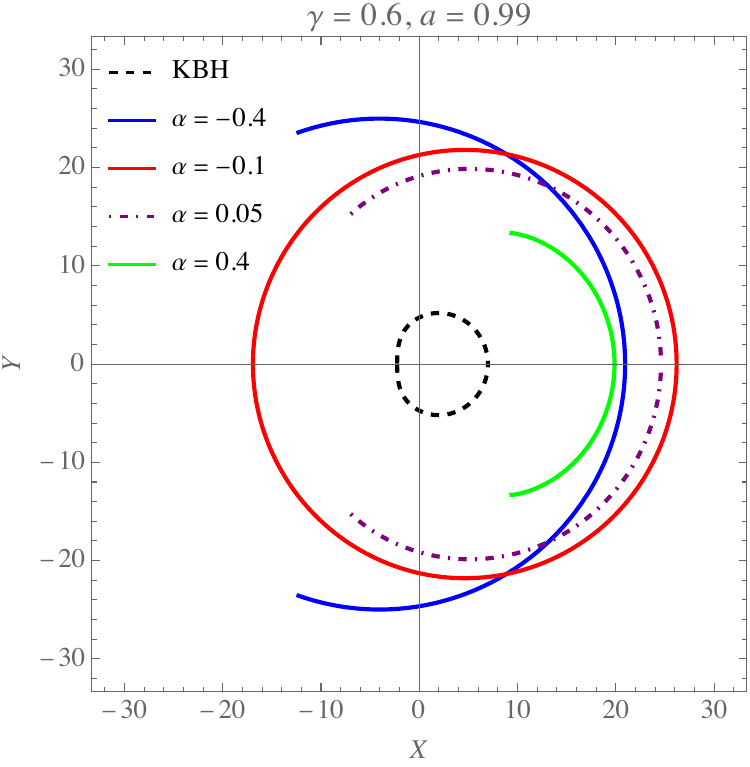} (h)
    \caption{The shape of the critical curves for the RMBH with $\theta_o=\pi/2$, corresponding to the following criteria: (a--d) fixed values for the $\alpha$ and $\gamma$ parameters with varying values of the spin parameter, (e--f) fixed values for the spin and $\alpha$ parameters with different curves corresponding to varying $\gamma$ parameter values, and (g--h) fixed values for the $\gamma$ and spin parameters with different curves representing varying $\alpha$ parameter values within both negative and positive domains. The unit of length along the axes is $M$.}
    \label{fig:shadow}
\end{figure}

Technically, in order to identify the shape of the black hole shadow, one might construct a celestial two-dimensional plane that spans the black hole's image in the eyes of a distant observer. For the case of the RMBH, since the spacetime is asymptotically finite, it is convenient to identify the celestial plane by means of the celestial coordinates \cite{Vazquez:2004}
\begin{eqnarray}
    && X = \lim_{r_o\rightarrow\infty}\left(-r_o^2\sin\theta_o \left.\frac{\ed\phi}{\ed r}\right|_{(r_o,\theta_o)}\right),\label{eq:X}\\
    && Y = \lim_{r_o\rightarrow\infty}\left(r_o^2 \left.\frac{\ed\theta}{\ed r}\right|_{(r_o,\theta_o)}\right),\label{eq:Y}
\end{eqnarray}
for an observer located at the point $(r_o, \theta_o)$, where $r_o$ is the observer's radial distance to the black hole and $\theta_o$ is the observer's inclination angle. Exploiting the equations of motion \eqref{eq:rdot}, \eqref{eq:thetadot}, and \eqref{eq:phidot}, one can derive the celestial coordinates for an observer at infinity. From the expressions \eqref{eq:A(r)} and \eqref{eq:B(r)}, it is evident that at infinity $A(r_o)=B(r_o)=1-\gamma$, and hence, using the definitions \eqref{eq:X} and \eqref{eq:Y}, the celestial coordinates are obtained as
\begin{eqnarray}
    && X = -\frac{\xi_p(r_p)}{\sin\theta_o}-\frac{a\gamma\sin\theta_o}{1-\gamma} ,\label{eq:X_1}\\
    && Y = \pm\sqrt{\eta_p(r_p)+a^2 \cos ^2\theta_o -\xi_p^2(r_p) \cot^2\theta_o} \,,\label{eq:Y_1}
\end{eqnarray}
which, as expected, incorporates the conical parameter $\gamma$. Obviously, for $\gamma=0$, the above coordinates resemble those for asymptotically flat spacetime (such as for KBH) as given in Ref.~\cite{Chandrasekhar:2002}. To demonstrate the critical curves, it is also convenient to locate the observer on the equatorial plane by setting $\theta_o=\pi/2$, and hence, the celestial coordinates that will be used for plotting the shadow boundaries take the form
\begin{eqnarray}
    && X = -{\xi_p(r_p)}-\frac{a\gamma}{1-\gamma} ,\label{eq:X_2}\\
    && Y = \pm\sqrt{\eta_p(r_p)} \,.\label{eq:Y_2}
\end{eqnarray}
Note that at each point on the critical curve in the celestial plane, one can consider a circle centered at the origin, whose radius $R_{c}$ can be calculated by using the expressions in Eqs. \eqref{eq:xip_2} and \eqref{eq:etap_2} in the celestial coordinates \eqref{eq:X_1} and \eqref{eq:Y_1}, which yields
\begin{eqnarray}
R_{c}^2 &=& X^2+Y^2\nonumber \\
&=& \frac{1}{2(1-\gamma)^2}\left[a^2\Bigl(3-2\gamma+\left(1-2\gamma\right)\cos(2\theta_o)\Bigr)+4\left(1-\gamma\right)K-\frac{8\left(1-\gamma\right)K'\Delta \Bigl(\Delta'-\left(1-\gamma\right)K'\Bigr)}{(\Delta')^2}
\right]_{r_p}.
    \label{eq:Rsh0}
\end{eqnarray}
Now, considering the critical constants in Eqs. \eqref{eq:xip_2} and \eqref{eq:etap_2}, we take $r_p$ as the curve parameter and generate the shadow boundaries in the $X$--$Y$ plane. In Fig.~\ref{fig:shadow}, several examples of the critical curves have been plotted for the RMBH with different spin parameters, while the parameters $\alpha$ and $\gamma$ are also considered to be varying in some examples.
As indicated by the diagrams, for a fixed value of $\alpha$, increasing the $\gamma$ parameter significantly enlarges the shadow compared to that of a KBH. Notably, for positive values of $\alpha$, the critical curves can resemble those of a Kerr naked singularity; however, here they correspond to RMBHs with either one or two horizons rather than to naked singularities. This is a distinctive feature of the RMBH spacetime, suggesting that high-spin RMBHs may manifest through unclosed photon rings. In these cases, photon regions do not form in the conventional sense, implying that the constants of motion $\xi$ and $\eta$, as given by Eqs.~\eqref{eq:xi_def} and \eqref{eq:eta_def}, do not permit a peak in the gravitational effective potential for photons approaching from all distances. Consequently, for certain black hole parameter values, the number of spherical photon orbits is reduced, with no solutions for $r_p$ under these conditions. In such scenarios, the majority of photons are either deflected toward an observer at infinity, forming alternative photon rings, or are deflected onto the event horizon and captured. Thus, the unclosed shadow boundaries correspond to a small subset of innermost unstable photon trajectories that reside on spherical orbits. Although previous studies have demonstrated that the shadows of KBHs and Kerr naked singularities may be indistinguishable to distant observers, it has also been shown that photon rings for black holes and naked singularities can exhibit observational differences \cite{Virbhadra:1998dy,virbhadra_gravitational_2002,virbhadra_time_2008,virbhadra_relativistic_2009,virbhadra_distortions_2022,tavlayan_instability_2024}. For the RMBH, however, as inferred from the critical curves, unclosed photon rings do not necessarily correspond to naked singularities. This is further evident from the shadow diagrams for positive values of $\alpha$, which confirm that no naked singularities are present in the spacetime.

\section{Shadow observables and constraints from the EHT }\label{sec:observables}

In this section, we begin by characterizing the shape of the black hole shadow through key observables, which we then apply to the RMBH to analyze the behavior of these observables specific to this black hole. Finally, we use these observables as tools to compare with EHT data to constrain the black hole parameters.

\begin{figure}[t]
    \centering
    \includegraphics[width=5.3 cm]{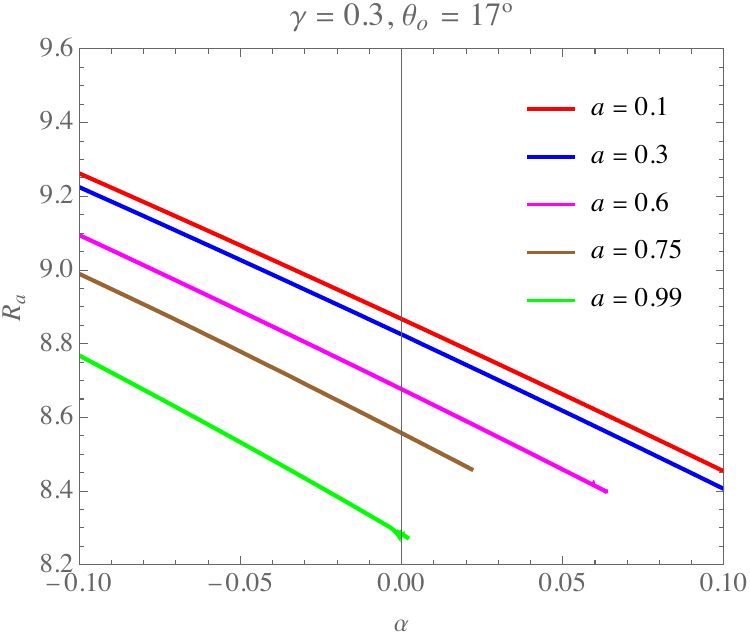} (a)\quad
    \includegraphics[width=5.3 cm]{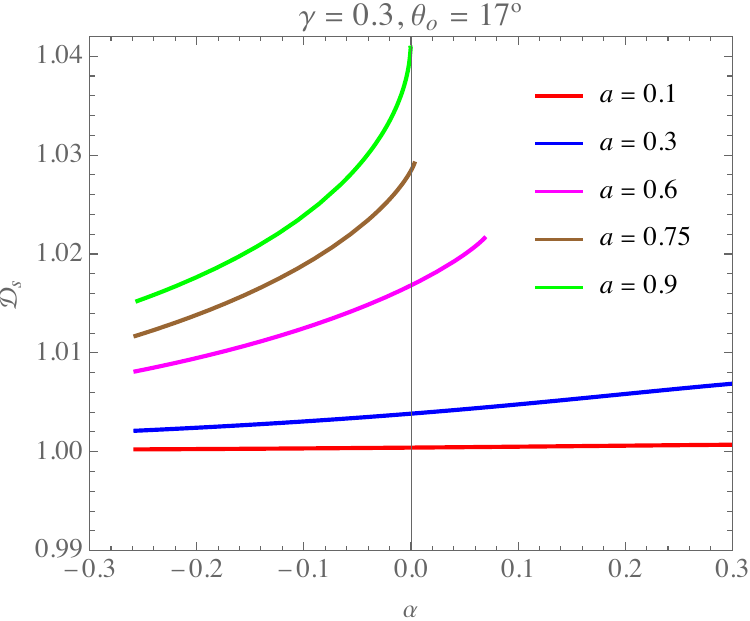} (b)\quad
    \includegraphics[width=5.3 cm]{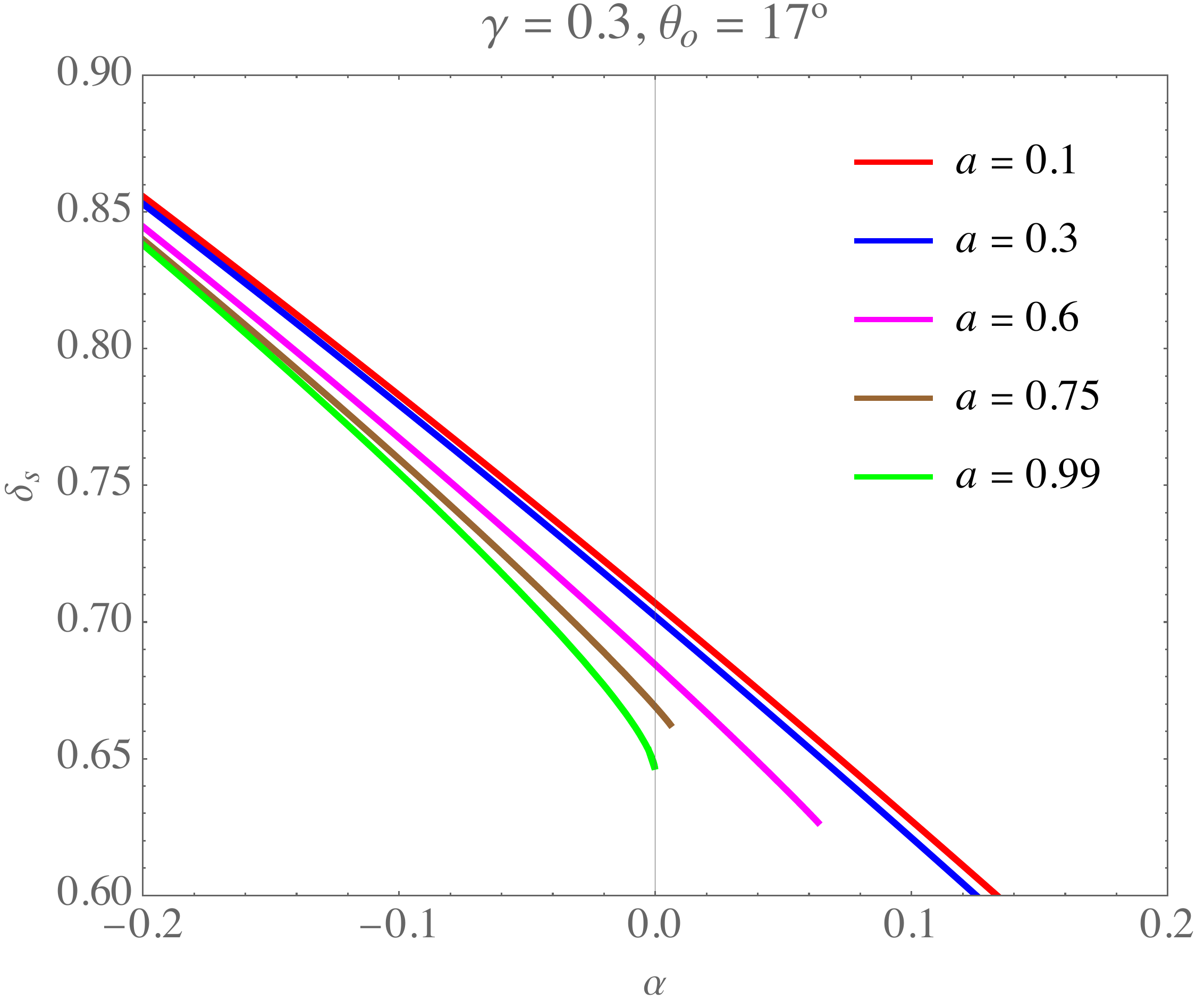} (c)
    \includegraphics[width=5.3 cm]{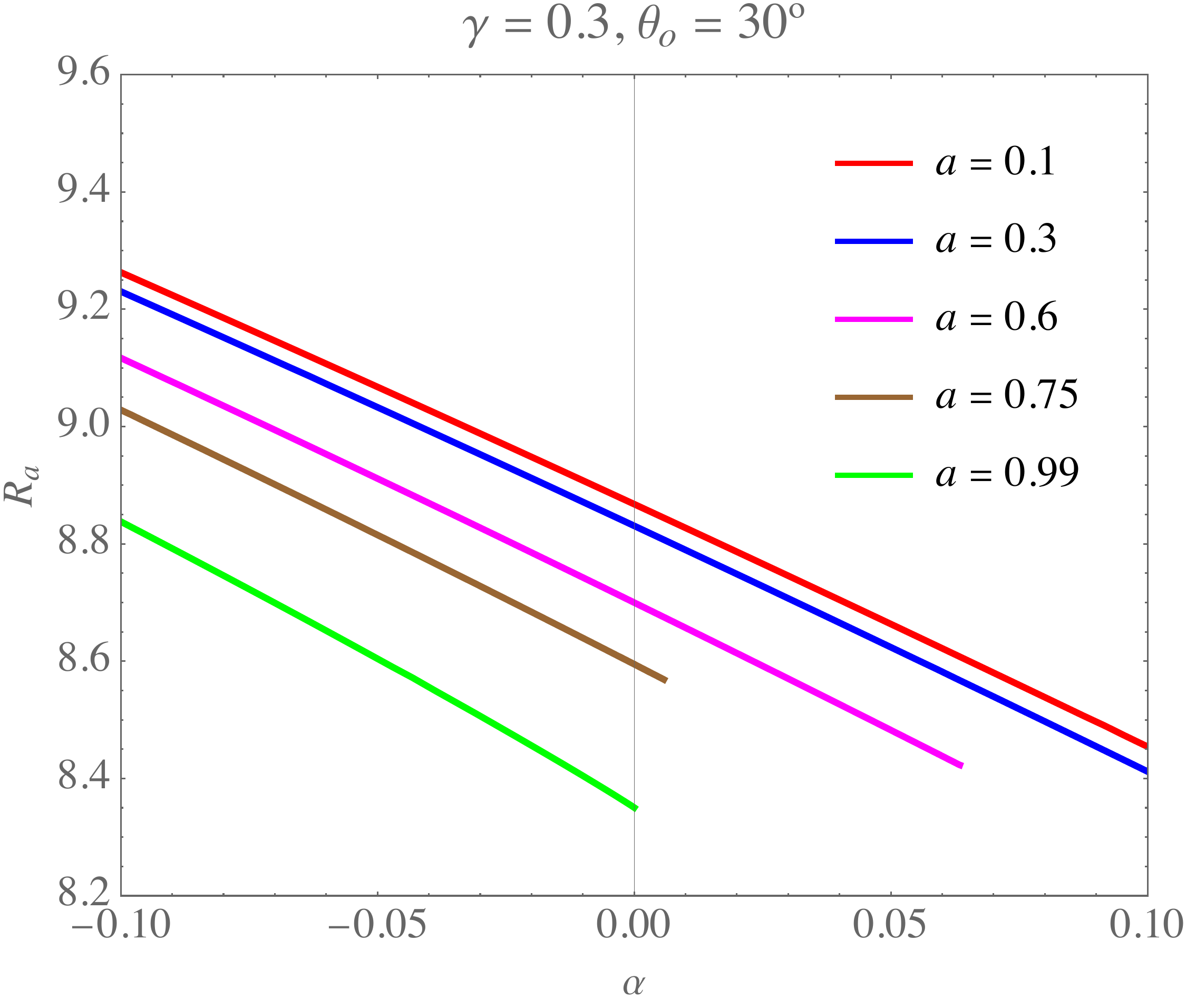} (d)\quad
    \includegraphics[width=5.3 cm]{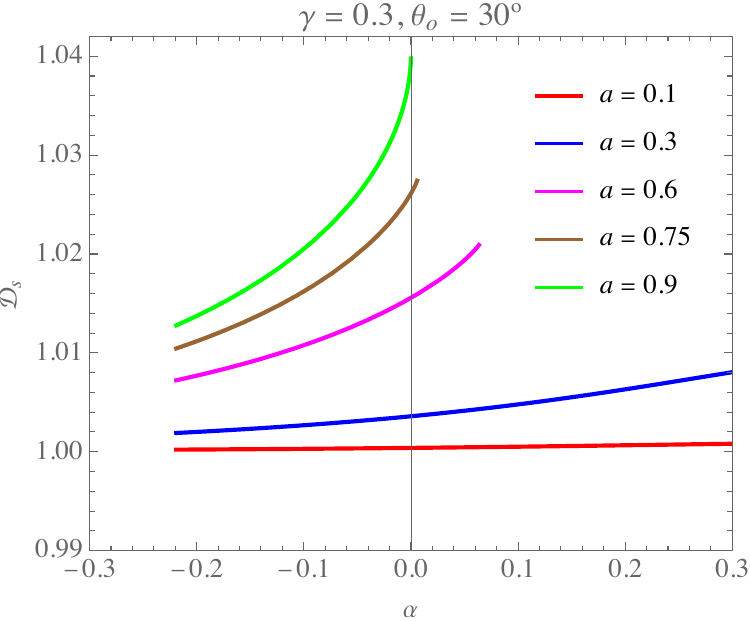} (e)\quad
    \includegraphics[width=5.3 cm]{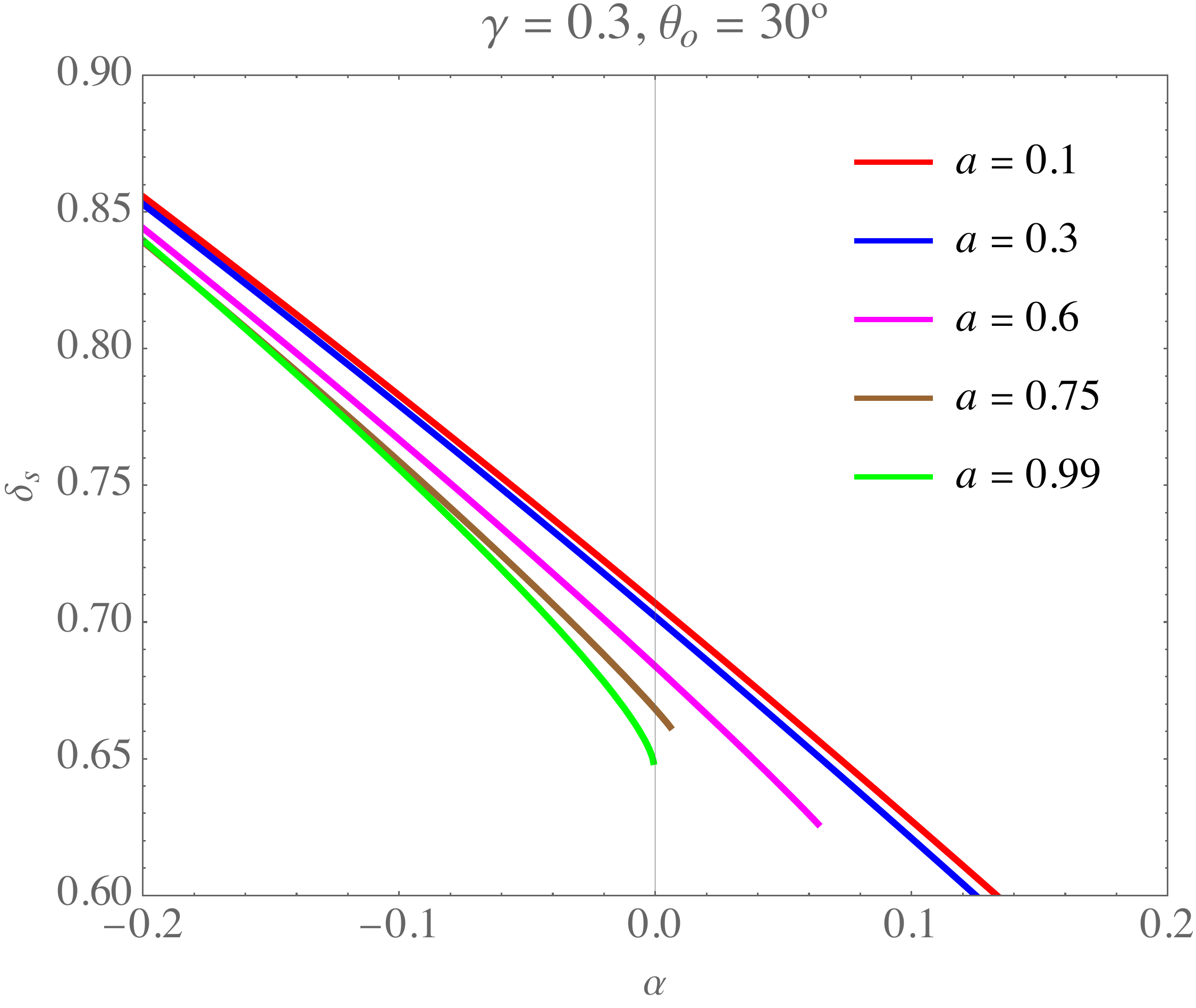} (f)
    \caption{The $\alpha$-profiles of $R_a$, $\mathscr{D}_s$, and $\delta_s$ are plotted for $\gamma = 0.3$ and different values of the spin parameter. The inclination angle for the panels in the first row is $17^\circ$, while for the second row, it is $30^\circ$. The unit of length for $R_a$ is $M$.}
    \label{fig:observables}
\end{figure}

Recently, numerous studies have explored the deformation and distortion of black hole shadows (see, for instance, Refs.~\cite{hioki_measurement_2009,johannsen_testing_2010,tsukamoto_constraining_2014,abdujabbarov_coordinate-independent_2015,amir_shapes_2016,tsupko_analytical_2017,ayzenberg_black_2018,afrin_parameter_2021,Kumar:2020a}). Following these approaches, the shadow can be characterized by the following features:

\begin{itemize}

    \item[(i)] The areal radius $R_a$, which quantifies the size of the black hole shadow, is defined as
    \begin{eqnarray}
        && R_a = \sqrt{\frac{\mathscr{A}_s}{\pi}},
        \label{eq:Ra}
    \end{eqnarray}
    in which $\mathscr{A}_s$ is the area enclosed by the critical curve, given by 
    \begin{equation}
    \mathscr{A}_s=2\int Y(r)\, \ed X(r)=2\int_{r_{p_-}}^{r_{p_+}}Y(r) X'(r)\,\ed r,
        \label{eq:As}
    \end{equation}
    where the radii of planar orbits $r_{p_\pm}$ are determined from the equation $Y^2(r_p)=0$.

    \item[(ii)] The shadow deformation $\mathscr{D}_s$, which describes the shadow’s asymmetry, is given by
    \begin{equation}
    \mathscr{D}_s = \frac{\delta Y}{\delta X} = \frac{Y_t-Y_b}{X_r-X_l}=\frac{2Y_t}{X_r-X_l},
        \label{eq:Ds}
    \end{equation}
    where the subscripts $t$, $b$, $l$, and $r$ denote the top, bottom, left, and right extremities of the shadow, as illustrated in Fig.~\ref{fig:schematic_shadow}, which depicts the geometric form of an oblate shadow. 
    \begin{figure}[t]
        \centering
        \includegraphics[width=6cm]{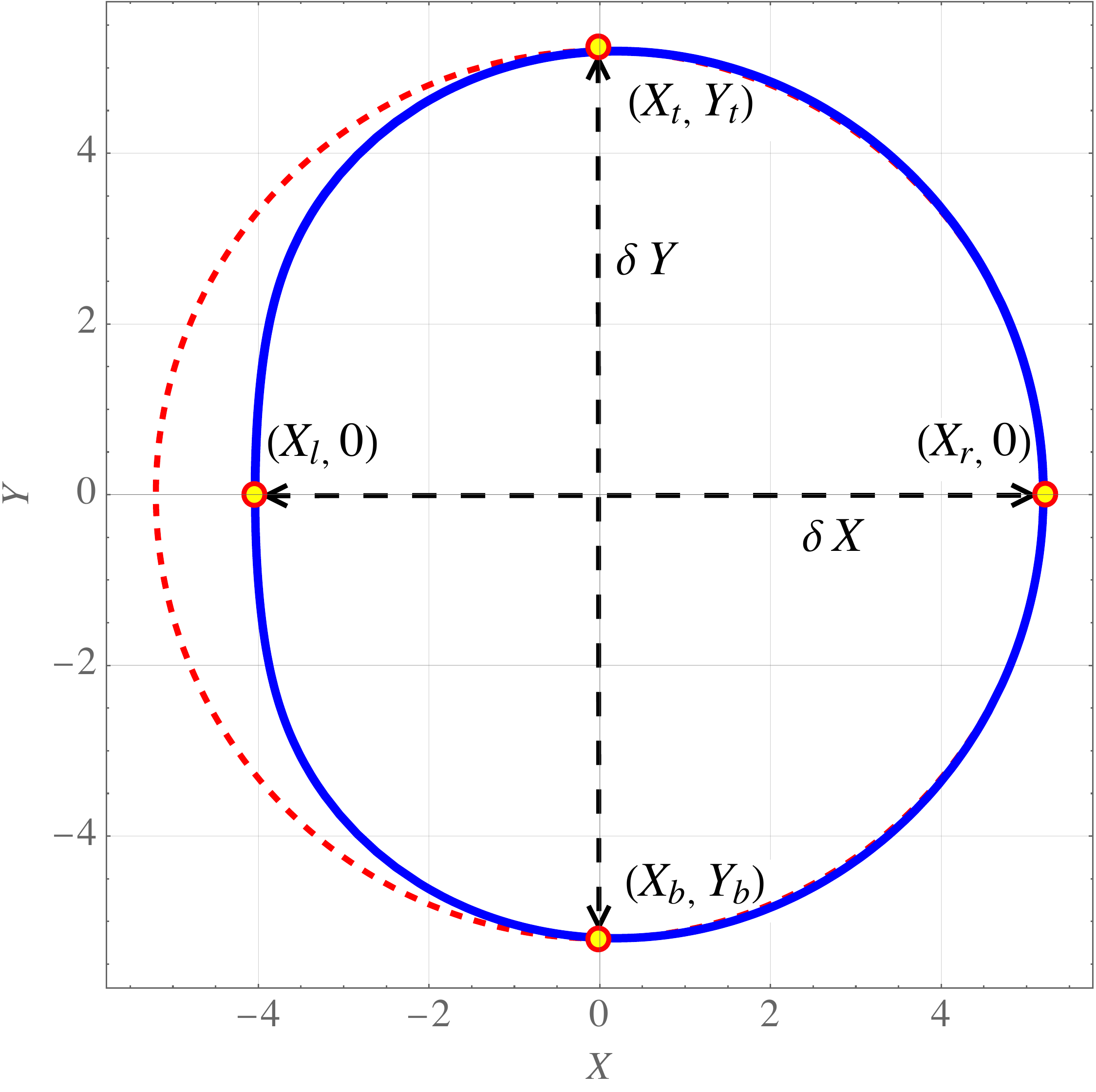}
        \caption{Schematic of an oblate shadow of a rotating black hole (solid blue curve) contrasted with a fully circular shadow (dashed red curve) used as reference, in $X$--$Y$ coordinates. The reference circle intersects the points $(X_b,Y_b)$, $(X_r,0)$, and $(X_t,Y_t)$.}
        \label{fig:schematic_shadow}
    \end{figure}
    For simplicity, we assume $Y_b=-Y_t$, justifying the factor of 2 in Eq.~\eqref{eq:As} due to the shadow’s reflectional symmetry along the $X$-axis.

    \item[(iii)] The fractional deviation parameter $\delta_s$, which measures the deviation of the shadow’s diameter from that of the SBH, is defined as
    \begin{equation}
        \delta_s=\frac{\bar{R}_{\rm{sh}}}{3\sqrt{3}}-1,
        \label{eq:deltas}
    \end{equation}
    where 
    \begin{equation}
    \bar{R}_{\rm{sh}}^2 =\frac{1}{2\pi} \int_0^{2\pi}\left[\left(X-X_c\right)^2+\left(Y-Y_c\right)^2\right]\ed\psi,
        \label{eq:barRsh}
    \end{equation}
    represents the average radius of the shadow, with $(X_c, Y_c=0)$ locating the geometric center of the shadow, where $X_c=(X_r+X_l)/2$. Additionally, 
    \begin{equation}
    \psi = \arctan\left(\frac{Y}{X-X_c}\right),
        \label{eq:psi}
    \end{equation}
    is the angular position of points on the shadow boundary relative to the center.
    
\end{itemize}
Note that, $X_r$ and $X_l$ are identified as $X(r_{p_\pm})$, while the top end $Y_t$ corresponds to $Y(r_t)$, where $r_t$ is determined from the condition ${\ed Y}/{\ed X} = 0$. Considering the celestial coordinates in Eqs.~\eqref{eq:X_1} and \eqref{eq:Y_1}, along with the expressions in \eqref{eq:xip_2} and \eqref{eq:etap_2}, the latter condition leads to the equation 
\begin{equation}
2\left(a^2s_o^2-\Delta\right)K'+\left[a^2\left(1-s_o^2\right)+K\right]\Delta'=0,
    \label{eq:rtEq}
\end{equation}
where $s_o \equiv \sin \theta_o$. In Fig.~\ref{fig:observables}, we present the behavior of the areal radius, the deformation, and the deviation as functions of the $\alpha$-parameter for two different inclination angles.
Note that the discontinuous curves in the diagrams correspond to the discontinuous photon rings (as illustrated in some examples in Fig. \ref{fig:shadow}). In these cases, the celestial coordinates \(X\) and \(Y\) cannot be computed as real quantities, typically because the light rays fail to orbit stably around the black hole (see the discussion in the previous section).  As observed from the diagrams in panels (a) and (d), for each fixed \(\gamma\)-parameter, the areal radius smoothly decreases within its continuous value domain as the \(\alpha\)-parameter increases. Furthermore, an increase in the spin parameter leads to a reduction in the areal radius, which is consistent with the expected behavior for classical Kerr-like black holes in GR.  Conversely, for a fixed \(\gamma\)-parameter, an increase in the \(\alpha\)-parameter results in a rise in the deformation, with a larger deformation for faster-rotating black holes. By comparing panels (b) and (e), we can observe that for the same parameter values, a higher inclination angle leads to an increase in the deformation.  Finally, as shown in panels (c) and (f), the deviation parameter is always positive, indicating that the average shadow radius of the RMBH is larger than that of the SBH and decreases rapidly as the \(\alpha\)-parameter increases. However, the positivity of \(\delta_s\) also implies that as the spin parameter increases, the deviation diminishes. This trend is evident in the diagrams for both inclination angles and is in agreement with the behavior of the areal radius demonstrated in panels (a) and (d).

\subsection{Constraints from the M87* observations}\label{subbsec:constraintsM87}

\begin{figure}[h]
    \centering
    \includegraphics[width=6cm]{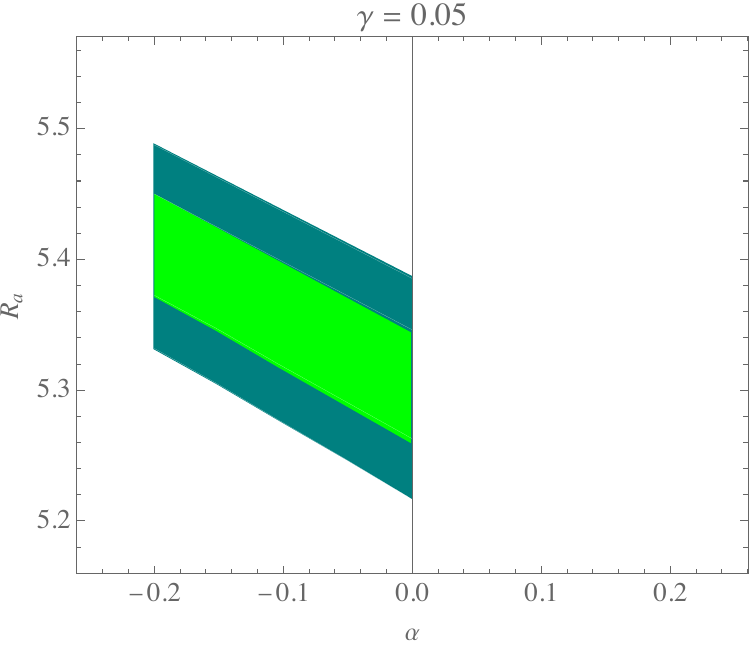} (a)\qquad
    \includegraphics[width=6cm]{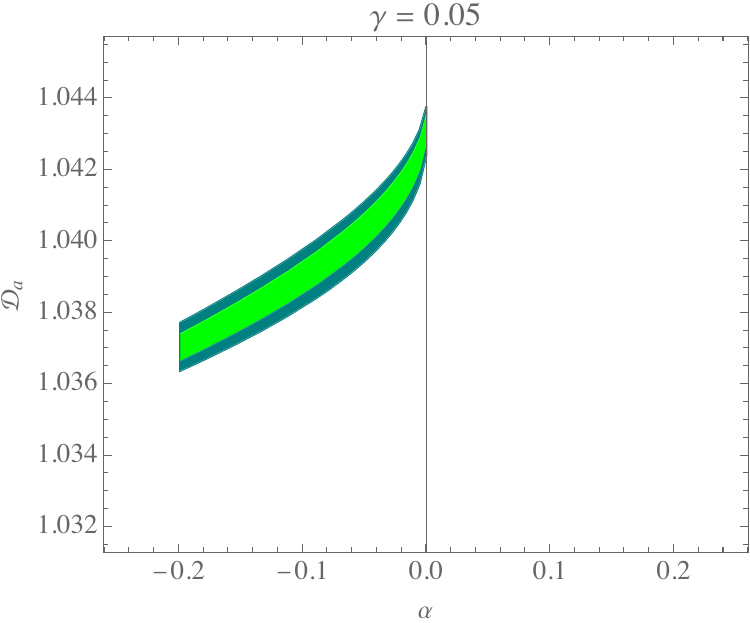} (b)
    \caption{Values of \(R_a\) and \(\mathscr{D}_s\) within the \(1\sigma\) (bright color) and \(2\sigma\) (dark color) uncertainties of the spin parameter of M87*, plotted for \(\gamma = 0.05\).
    The unit of length along the \(R_a\) axis is \(M\).}
    \label{fig:constnats_M87}
\end{figure}

The EHT observational results from the shadow images of M87* revealed the areal radius and deformation of the shadow, respectively, as \(4.31M \leq R_a \leq 6.08M\) and \(1 \leq \mathcal{D}_s \leq 1.33\) \cite{Akiyama:2019_L1}. Furthermore, it has been shown that the spin parameter of M87* is approximately \(a = (0.9 \pm 0.05) M\) \cite{tamburini_measurement_2020}. Following the discussion in Ref. \cite{daly_new_2023}, we assume \(\theta_o = 17^\circ\). Considering the spin parameter of M87* mentioned above, we observe from panels (a) and (d) of Fig. \ref{fig:observables} that the values of \(R_a\) for the corresponding curve do not lie within the range observed for M87*. This suggests that the \(\gamma\)-parameter should be smaller. In the left panel of Fig. \ref{fig:constnats_M87}, we have plotted the areal radius and the deformation as functions of \(\alpha\), for the spin parameter of M87* within its \(1\sigma\) and \(2\sigma\) uncertainties, where the \(\gamma\)-parameter has been fixed to \(0.05\). 

As observed from the behavior of \(R_a\), the value \(\gamma = 0.05\) places the domain of confidence within an acceptable range regarding the M87* observations. However, considering the range of \(R_a\) covered within the \(2\sigma\) uncertainty, it can be concluded that this parameter will still lie within the confident domain of M87* observations if the chosen value of \(\gamma\) is doubled. Thus, we infer that the \(\gamma\)-parameter performs best within the range \(0 \leq \gamma \lesssim 0.1\). Furthermore, the physically reliable range of \(\alpha\) within these data is \( -0.2 \lesssim \alpha \leq 0\), which can also be regarded as its observational constraint. Therefore, the negative branch of \(\alpha\) holds more astrophysical significance. These constraints have been tested for the deformation in Fig. \ref{fig:constnats_M87}(b). Note that the range of \(\mathscr{D}_s\) in diagrams (b,e) of Fig. \ref{fig:observables} for \(\gamma = 0.3\) is already within the acceptable ranges, particularly for \(a = 0.9M\). However, we see that for \(\gamma = 0.05\), the results are still acceptable. Hence, we can confidently adopt the new domain for \(\gamma\) as asserted above, since it respects the validity of \(R_a\) within the observational data.

\subsection{Constraints from the Sgr A* observations}\label{subsec:constraintsSgrA}

\begin{figure}[h]
    \centering
    \includegraphics[width=6cm]{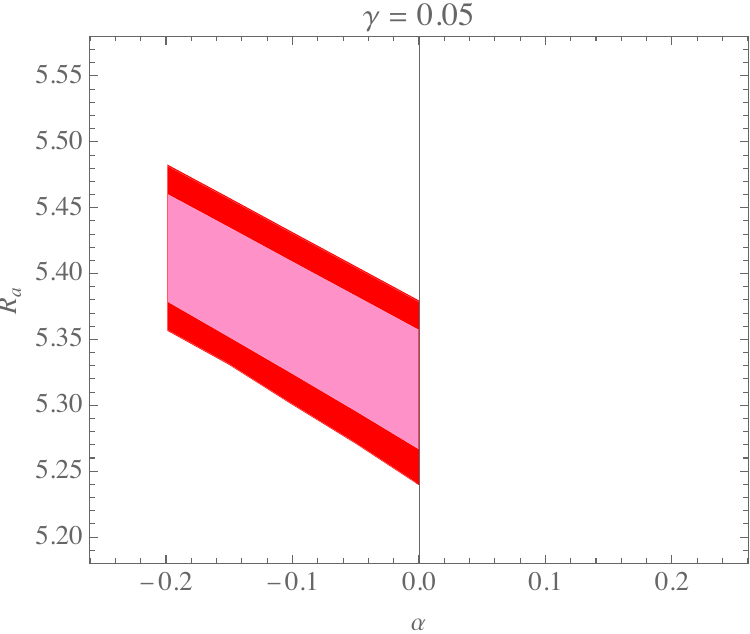} (a)\qquad
    \includegraphics[width=6cm]{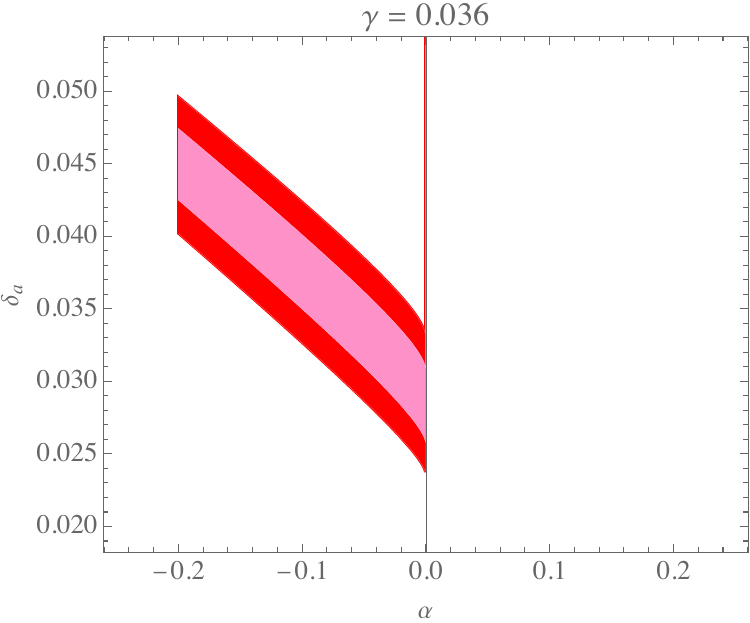} (b)
    \caption{Values of \(R_a\) plotted for \(\gamma = 0.05\), and \(\delta_s\) plotted for \(\gamma = 0.036\), within the \(1\sigma\) (bright color) and \(2\sigma\) (dark color) uncertainties of the spin parameter of Sgr A*. The unit of length along the \(R_a\) axis is \(M\).}
    \label{fig:constnats_SgrA}
\end{figure}

The parameters of Sgr A* observed by the EHT are specified by the Very Large Telescope Interferometer (VLTI) and Keck observatories. These observations have constrained the areal radius and the fractional deviation peculiar to Sgr A* as \(4.5M \leq R_a \leq 5.5M\) and \(-0.17 \leq \delta_s \leq 0.01\) by VLTI, and \(4.3M \leq R_a \leq 5.3M\) and \(-0.14 \leq \delta_s \leq 0.05\) by Keck \cite{Akiyama:2022}. Furthermore, the inclination angle determined by the EHT was approximately \(\theta_o < 50^\circ\) for two classes of models used in simulations. For this purpose, two different spin parameters, \(a = 0.5M\) and \(a = 0.94M\), were applied to satisfy a series of EHT constraints in the general relativistic magnetohydrodynamic (GRMHD) simulations. On the other hand, recent analysis of Sgr A* suggests that its spin should be \(a = (0.9 \pm 0.06)M\) \cite{daly_new_2023}. This value is taken into account in this subsection to constrain the parameters of the RMBH. Additionally, a reliable inclination of \(30^\circ\) is assumed. In Fig. \ref{fig:constnats_SgrA}, we have plotted the behavior of \(R_a\) and \(\delta_s\) for the RMBH, considering the spin parameter of Sgr A* within its \(1\sigma\) and \(2\sigma\) uncertainties.
As inferred from the left panel, the values of the areal radius lie within the observational data derived from VLTI and Keck, with \(\gamma = 0.05\) as the upper limit. This indicates that larger values of \(\gamma\) would violate the observational constraints. Therefore, this suggests that, while the \(\alpha\)-parameter can still respect the domain \(-0.2 \lesssim \alpha \leq 0\), the \(\gamma\)-parameter should be less than 0.05. This is also consistent with the behavior of \(\delta_s\) within Sgr A* observations, as for the chosen value of \(\gamma = 0.036\), the fractional deviation becomes \(\delta_s \approx 0.05\), which is the upper limit of this parameter according to the Keck observations. Hence, to be consistent with the Sgr A* observations, the \(\gamma\)-parameter should lie within the domain \(0 \leq \gamma \lesssim 0.036\). Notably, this range just covers a portion of the acceptable values constrained by the M87* observational data. Thus, we can infer that the Sgr A* observations impose more stringent constraints on the parameters of the RMBH.

\begin{figure}[h]
    \centering
    \includegraphics[width=6cm]{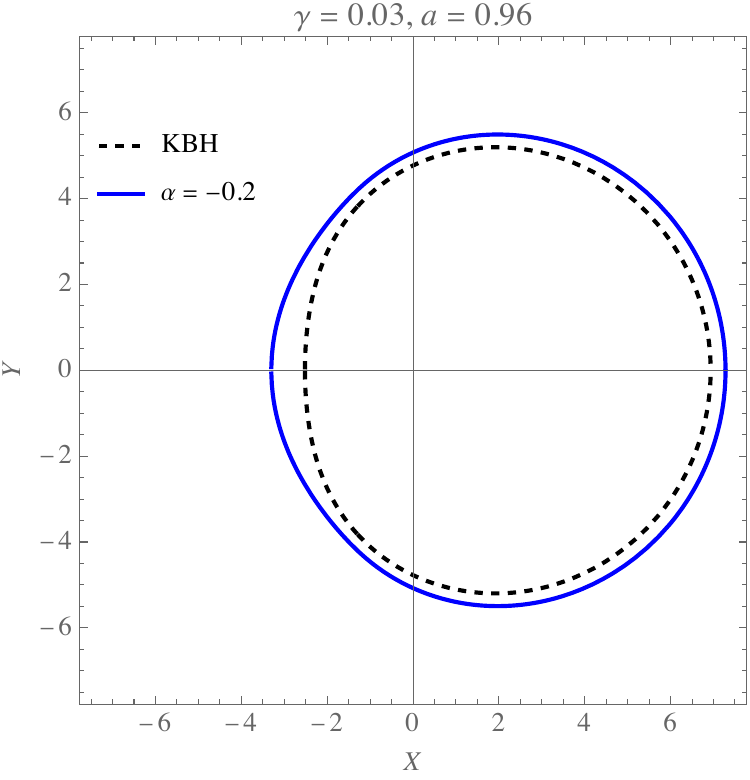} 
    \caption{The shadow boundary of the RMBH plotted for \(\alpha = -0.2\), \(\gamma = 0.03\), \(a = 0.96\), and \(\theta_o = \pi/2\), compared with the shadow of a KBH for the same spin parameter. The unit of length along the axes is \(M\).}
    \label{fig:shadow_const}
\end{figure}

Based on these constraints, in Fig. \ref{fig:shadow_const}, we have plotted the shadow of the RMBH, considering the values for the \(\alpha\) and \(\gamma\)-parameters that are in agreement with both the M87* and Sgr A* observations, as discussed above.
As we can see, within the observational constraints, the shadow radius of the RMBH is slightly larger than that of a KBH. Furthermore, in contrast to the oblateness of the shadow of a fast rotating KBH, the shadow of a fast rotating RMBH exhibits sharpness. This is an important feature of the weakly coupled GMC in the spacetime of the RMBH, which manifests itself in the peculiar shape of the shadow.

\section{The energy emission rate}\label{sec:energy}

\begin{figure}[h]
    \centering
    \includegraphics[width=5.3cm]{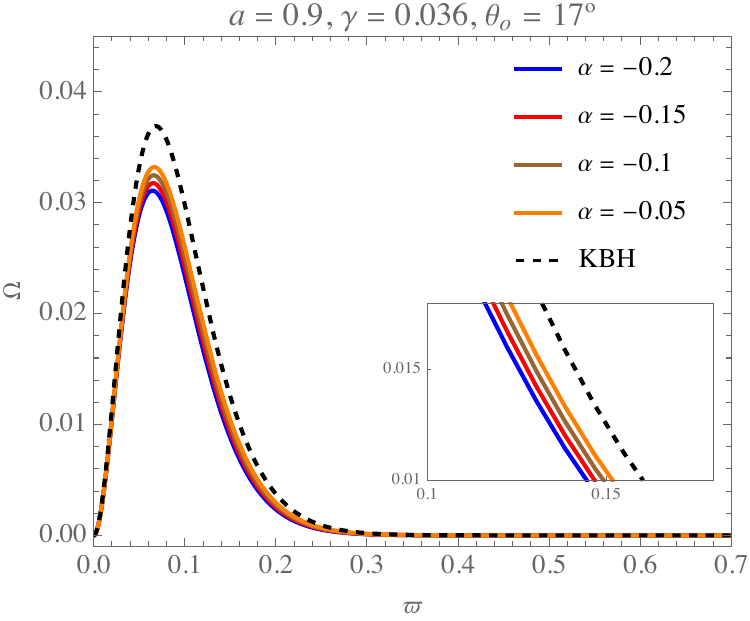} (a)\quad
    \includegraphics[width=5.3cm]{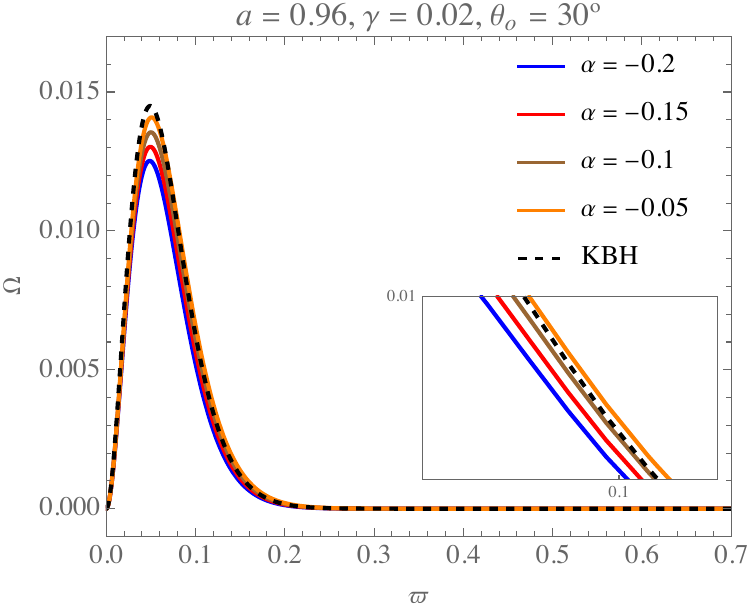} (b)\quad
    \includegraphics[width=5.3cm]{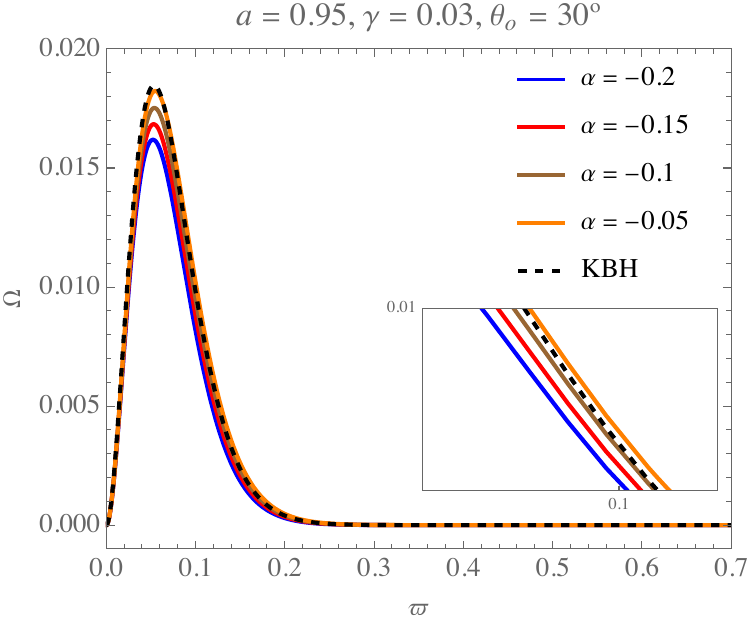} (c)
    \caption{The profiles of the energy emission rate with respect to changes in the frequency \(\varpi\), for three different cases of the spin parameter, the \(\alpha\)-parameter, and the \(\gamma\)-parameter within the accepted observational constraints, and two different inclinations, compared to the energy emission rate for the KBH for the same spin parameters.}
    \label{fig:Eomega}
\end{figure}

From the quantum mechanical perspective, on the black hole's event horizon, particles can be created and annihilated, and those with positive energy can escape the event horizon by tunneling. This process sets black holes into radiation, which may eventually cause them to evaporate. Black hole evaporation occurs due to Hawking radiation, a quantum effect in which black holes emit thermal radiation, gradually losing mass and energy over time until they eventually vanish \cite{Hawking:1974rv}. At high energy levels, Hawking radiation is typically emitted within a finite cross-sectional area, denoted as \(\sigma_l\). For distant observers positioned far from the black hole, this cross-section gradually approaches the shadow cast by the black hole \cite{belhaj_deflection_2020, wei_observing_2013}. It has been shown that \(\sigma_l\) is directly linked to the area of the photon ring and can be mathematically represented as \cite{wei_observing_2013, decanini_fine_2011, li_shadow_2020}
\begin{equation}
\sigma_l \approx \pi \bar{R}_{\rm{sh}}^2.
    \label{eq:sigmal}
\end{equation}
Accordingly, the energy emission rate of the black hole can be expressed as
\begin{equation}
\Omega \equiv \frac{\ed^2 E(\varpi)}{\ed\varpi \, \ed t} = \frac{2\pi^2 \sigma_l}{e^{\varpi / T_\mathrm{H}^+} - 1} \varpi^3 \approx \frac{2\pi^3 \bar{R}_{\rm{sh}}^2 \varpi^3}{e^{\varpi / T_\mathrm{H}^+} - 1},
    \label{eq:emissionrate}
\end{equation}
where \(\varpi\) is the emission frequency, and \(T_\mathrm{H}^+ = {\kappa}/{2\pi}\) is the Hawking temperature at the event horizon, where
\begin{equation}
\kappa = \left. \frac{\Delta'(r)}{2\left(a^2 + r^2\right)} \right|_{r_+},
    \label{eq:kappa}
\end{equation}
is the surface gravity at the event horizon. It is straightforward to verify that for zero spin parameter (i.e. \(a = 0\)), this quantity reduces to \(\kappa = A'(r_+)/2\), which is the surface gravity of the event horizon for static black holes. In Fig. \ref{fig:Eomega}, some examples of the behavior of \(\Omega\) versus changes in the frequency \(\varpi\) are plotted for the RMBH, with black hole parameters taken within the observationally accepted domains, as identified in the previous section.
As observed from the diagrams, the larger the \(\gamma\)-parameter, the higher the energy emission rate, for all values of the coupling constant \(\alpha\). Hence, the GMC results in a faster evaporating black hole. It is also observed that faster rotating black holes emit less energy. Consequently, for higher spin parameters, the profiles of the RMBH resemble those of the KBH. Furthermore, as the observer's inclination increases, the energy emission rate decreases. According to the diagrams, we can infer that the largest emission rate for the RMBH corresponds to the lower limit of the spin parameter, the upper limit of the coupling constant, and the upper limit of the GMC. However, note that such cases exhibit significantly lower energy emission rates than the KBH.

\section{Summary and conclusions}\label{sec:conclusion}

In This study we analyzed the shadow properties of a rotating black hole with a weakly coupled GMC, i.e. the RMBH, which is derived by applying MNJA from the seed static spacetime. {We showed that the GMC, by creating a deficit angle, modifies the black hole’s causal structure, photon regions, and shadow observables, while general relativity remains the background theory.} Our findings reveal that the combined influence of the GMC and rotation produces distinct shadow features, which could potentially differentiate the RMBH from the Kerr-like black holes of GR. Our shadow analysis demonstrated how both the GMC and the weak coupling constant modify the size, circularity, and shape of the shadow, resulting in deviations that may be detectable in astrophysical observations. The investigation of critical curves in the shadow showed that, for specific values of the GMC and coupling parameters, certain photon trajectories remain open, especially under high-spin conditions. These open photon rings indicate incomplete spherical photon orbits, forming unique shadow boundaries for rotating black holes with GMC. Such characteristics may provide observable signatures that help distinguish the RMBH from the KBH. Moreover, using shadow observables, including the areal radius \( R_a \) and the shadow deformation parameter \( \mathscr{D}_s \), we constrained the GMC parameter \( \gamma \) and weak coupling constant \( \alpha \) by comparing theoretical shadow predictions with observational data from the EHT for M87* and Sgr A*. This analysis restricted the GMC parameter to \( 0 \leq \gamma \lesssim 0.036 \) and the weak coupling constant to \( -0.2 \lesssim \alpha \leq 0 \). These constraints provide important experimental limits on the parameter space for black holes with GMC, offering insights into the symmetry-breaking scale associated with global monopoles in astrophysical scenarios. We further examined the black hole's energy emission rate, observing that higher values of the GMC correspond to increased emission rates, suggesting a faster evaporation process. Our analysis indicates that emission rates are inversely related to the spin parameter, with higher rotation leading to a reduced emission rate. Additionally, observers at larger inclination angles experience decreased emission rates, suggesting that inclination angle and rotation significantly influence the black hole’s energy profile. In conclusion, our study underscores the potential of black hole shadow observations as a tool for testing nonminimal coupling theories and detecting topological defects like global monopoles. {By constraining the GMC and coupling parameter, our results also provide a potential phenomenological test for nonminimal gravity.} Future research may extend this analysis by exploring additional shadow observables, such as asymmetry or lensing rings, or by examining the behavior of GMC in stronger coupling regimes. Such investigations could further clarify the role of symmetry-breaking mechanisms and topological defects in shaping black hole spacetimes, providing valuable insights into modified gravity and early-universe physics.

\begin{acknowledgments}
This work has been supported by Universidad Central de Chile through project No. PDUCEN20240008.  
\end{acknowledgments}

\section*{Funding}
This research received no external funding.

\section*{Data Availability Statement}
There is no data associated with this study. 

\section*{Conflicts of Interest}
The authors declare no conflict of interest.

\bibliographystyle{ieeetr}
\bibliography{biblio_v1}

\end{document}